\RequirePackage{fix-cm}
\documentclass[natbib,smallextended]{svjour3}       
\smartqed  
\usepackage{graphicx}
 \journalname{my journal}
\begin{document}

\title{Electric Current Circuits in Astrophysics
}


\author{Jan~Kuijpers         \and
        Harald~U.~Frey  \and  Lyndsay~Fletcher 
}


\institute{L. Fletcher \at
              SUPA School of Physics and Astronomy\\
              University of Glasgow\\
              Glasgow,  G12 8QQ\\
              UK\\
	     Tel.: +44 141 330 5311\\              
	     Fax: +44 141 330 8600\\
              \email{lyndsay.fletcher@glasgow.ac.uk}           
           \and
           H.U. Frey \at
              Space Sciences Laboratory\\
              7 Gauss Way\\
              Berkeley, CA 94720-7450\\
              USA\\
              Tel.: +1-510-643-3323\\
              Fax: +1-510-643-2624\\
              \email{hfrey@ssl.berkeley.edu}
            \and
            J.Kuijpers\\
            Department of Astrophysics/ IMAPP\\
            Radboud University Nijmegen\\
            P.O. Box 9010\\
            6500 GL Nijmegen\\
            The Netherlands\\
            Tel.: +31 6 269 747 83\\
            Fax: +31 24 36 52807\\
            \email{kuijpers@astro.ru.nl}
}

\date{Received: date / Accepted: date}

\maketitle

\begin{abstract}
Cosmic magnetic structures have in common that they are anchored in a dynamo, that an external driver converts kinetic energy into internal magnetic energy, that this magnetic energy is  transported as Poynting flux across the magnetically dominated structure, and that the magnetic energy is released in the form of particle acceleration, heating, bulk motion, MHD waves, and radiation. The investigation of the electric current system is particularly illuminating as to the course of events and the physics involved. We demonstrate this for the radio pulsar wind, the solar flare, and terrestrial magnetic storms.

\end{abstract}

\section{Introduction}
\label{intro}
Magnetic field structures in the cosmos occur on many scales, spanning a range of over 15 decades in spatial dimension, from extragalactic winds and jets down to the terrestrial magnetosphere. Yet in all these objects the properties of magnetic fields lead to a very similar, multi-scale, spatial and temporal structure. Magnetic fields originate in electric currents, as described by Maxwell's equations. They have energy, pressure, and tension, as quantified by their energy-momentum (stress) tensor. They exert a force on ionized matter, as expressed by the Lorentz force. Finally, their equivalent mass is small when compared to the mass-energy of ambient matter. 

These basic properties lead to a common appearance in a variety of objects which can be understood as follows. Since there are no magnetic charges, magnetic fields are not neutralized and they extend over large stretches of space and time. The Lorentz force which keeps ionized matter and magnetic fields together allows gravitation to anchor a magnetic field in a condensed ionized object. Since there gas and dynamic pressures dominate over the magnetic pressure, magnetic fields tend to be amplified by differential motions such as occurs in a  stellar convection zone, a differentially rotating accretion disk, binary motion, or  a stellar wind around a planet. Expressed in circuit language, a voltage source is set up by fluid motions which drives a current and forms a dynamo in which magnetic field is amplified at the expense of kinetic energy. Next, the small equivalent mass of the magnetic fields makes them buoyant, and as ionized matter slides easily along magnetic fields, they pop up out of the dense dynamo into their dilute environs. Whereas this central domain is dominated by fluid pressure the environs are dilute so that there the magnetic pressure dominates. The tension of the magnetic field then allows for transport of angular momentum from the central body along the field outward into a magnetized atmosphere such as a corona or magnetosphere. As more and more magnetic flux rises into the corona, and/or differential motions at the foot-points of magnetic structures continue to send electric currents and associated Poynting flux into the corona the geometries of these nearly force-free structures adapt, expand and generally lead to the appearance of thin sheets of concentrated electric currents. The same process occurs in the terrestrial magnetosphere but now the Poynting flux is going inward toward the central body. Finally, the magnetic structure in the envelope becomes unstable and the deposited energy is released in a process equivalent to an  electromotor or Joule heating but now in a multitude of small `non-force-free' regions created by the currents themselves, often together in   explosions, such as storms, (nano-)flares, ejected plasmoids, jets, but also in a more steady fashion, such as super-fast winds. 

The description of the formation and dynamics of complicated magnetic structures in terms of simplified electric current circuits in a variety of objects elucidates the fundamental physics by demanding consistency, and distinguishing cause and effect. Also, it allows for a unified answer to a number of pertinent questions:
\begin{itemize}
\item {\it Current Distribution}: what is the voltage source, how does the current close, which domains can be considered frozen-in, where (and when) is the effective resistivity located? 
\item {\it Angular Momentum Transport}: where are the balancing (decelerating and accelerating) torques?
\item {\it Energy Transport}: what is the relative importance of Poynting flux versus kinetic energy flux?
\item {\it Energy Conversion}: what is the nature of the effective resistivity (Lorentz force, reconnection, shocks), and what is the energy partitioning, i.e. the relative importance of gas heating, particle acceleration, bulk flow, and MHD waves?
\end{itemize} 
For this review we have chosen to zoom in on the magnetosphere of the {\it radio pulsar}, on the {\it solar flare}, and on {\it terrestrial aurora and magnetic storms}. We will point out parallels and similarities in the dynamics of the multi-scale magnetic structures by considering the relevant electric circuits. Many of the same questions (and answers) that are addressed below are relevant to other objects as well, such as extragalactic jets, gamma ray bursts, spinning black holes, and planetary magnetospheres.

Equations will be given in Gaussian units throughout to allow simple comparisons to be made while numerical values are in a variety of units reflecting their usage in the fields they come from.

\section{Electric circuit of the pulsar wind}
\label{psr}
The pulsar wind presents an important class of Poynting flux-dominated outflows. These are dominated by electromagnetic rather than kinetic energy. Here we want to study the closed electric current system, which involves regions both very near to the pulsar and very far away in the pulsar wind, a problem for which no general agreement exists as to its solution. The relative importance of the magnetic and kinetic energy flows is conveniently written as the magnetization $\sigma$:
\begin{equation}
\sigma\equiv \frac{\rm Poynting \;flux}{\rm kinetic\; energy \;flux}=\left\{
\frac{{B^2_{tor}/{4\pi}}}{\rho v^2_{pol}/2}\;,\;\frac{B^2_{tor}/4\pi}{\Gamma\rho c^2}\right\}>1,
\label{eq:1}
\end{equation}
where we have included the contribution only from the toroidal  (i.e. transverse to the radial direction) magnetic field $B_{tor}$ and neglected the poloidal (i.e. in the meridional plane) magnetic field $B_{pol}$ since the latter falls off faster with distance than the former in a steady wind. $v_{pol}$ is the poloidal component of the wind velocity, $\rho$ the wind matter density in the observer frame, and $\Gamma$ the Lorentz factor of the wind. The first term inside curly brackets applies to non-relativistic and the second to relativistic winds. Further, it is assumed that either the ideal MHD condition applies in the wind: 
\begin{equation}
\vec E = -\vec{v} \times \vec B/c,
\label{eq:2a}
\end{equation}
or that the wind is nearly a vacuum outflow in which case the condition 
\begin{equation}
\vec E= -\vec c \times \vec B/c
\label{eq:2b}
\end{equation}
takes over smoothly from the ideal MHD condition.  Here $\vec B$ is the magnetic induction (magnetic field), $\vec v$ the fluid velocity, and $c$ the speed of light.

\begin{figure}
\includegraphics[width=0.5\textwidth]{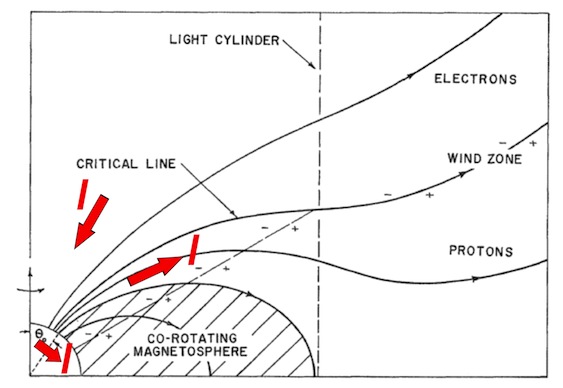}
\includegraphics[width=0.5\textwidth]{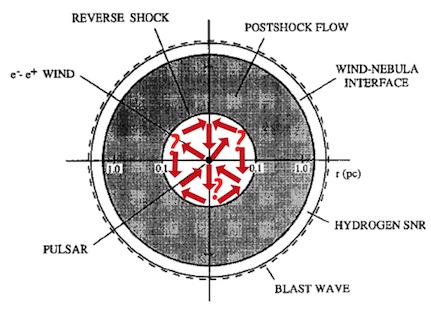}
\caption{{\bf Left}: The Goldreich-Julian picture of the pulsar magnetosphere for an aligned magnetic rotator \protect\citep{Goldreich:1969fk}. The potential difference across the polar cap drives a current $ I$, which enters the star along polar field lines, crosses the stellar surface layers, leaves the star along field lines on the other side of the `critical' field line, and is supposed to close somewhere in the wind zone. Note that the incoming/ outgoing current is carried here by respectively outgoing electrons/ outgoing protons, but could equally well be carried by a surplus of outgoing electrons/ outgoing positrons. Adaptation from \protect\citet{Goldreich:1969fk}, reproduced by permission of the AAS.
{\bf Right}: Sketch of the closure of the current system in the Crab wind, which has to take place before the reverse shock at about 0.1 pc.}
\label{fig:1}      
\end{figure}

The rotating magnetized star which forms the pulsar is a so-called `unipolar inductor' \citep{Goldreich:1969fk}. The rotation of the magnetized conductor creates a potential drop across the moving field lines from the magnetic pole towards the equator (see Figure \ref{fig:1}, Left). This voltage drop appears along the 
field lines between the star and infinity. The reaction of the star to this strong field-aligned potential drop is that, under certain conditions, an electric current is set up (see Figure \ref{fig:1}, Left). Charged particles are drawn out of the crust and accelerated to such high energies that  a dense wind of electron-positron pairs leaves the star and provides for the electric currents. The magnitude of the electric current density near the stellar surface is expected to be of order 
\begin{equation}
j_{GJ}(r_\star)\equiv \tau_{GJ}(r_\star)c,
\label{eq:2}
\end{equation}
where the Goldreich-Julian (GJ) charge density is defined as
\begin{equation}
\tau_{GJ}(r)\equiv -\frac{\vec \Omega_\star \cdot \vec B_0(\vec r)}{2\pi c}.
\label{eq:2c}
\end{equation} 
$\vec B_0(\vec r)$ is the background magnetic field at position $\vec r$, $\vec \Omega_\star$ is the stellar rotation vector, and $\vec r_\star$ the stellar radius. The GJ charge density just provides for a purely transverse electric field and a corresponding $\vec E \times \vec B$-drift which causes the (ideal) plasma to co-rotate with the star at the angular speed $\vec\Omega_\star$ (\ref{eq:2a}). As a result, when the charge density is equal to the local GJ density everywhere, the parallel electric field vanishes. This is the situation on the `closed' field lines which are located near the star. 

On the open field lines  the speed of the charges is assumed to be the speed of light since the wind is expected to be relativistic from the beginning (\ref{eq:2}). Things are complicated here because a strictly steady state pair creation is not possible. It is clear that the {\it parallel} electric field momentarily vanishes as soon as the charge density reaches the value $\tau_{GJ}$. However, to produce this GJ-density one needs a very strong parallel electric field to exist. Actually, the strong time-variations within a single radio pulse are believed to mirror the temporal process of pair creation. For our purpose we assume a steady-state relativistic wind to exist in the average sense. This is the reason for the appearance of the GJ density in (\ref{eq:2}). The current (and the much denser wind) only exist on the so-called `open' field lines, the field lines which connect to infinity, (see Figure \ref{fig:1}, Left).  For a steady current the incoming current (poleward for the aligned rotator of Figure \ref{fig:1}, Left) must be balanced by the outgoing current (equator-ward). This defines a critical field line in the wind (Figure \ref{fig:1}, Left) separating the two parts of the current. Note, however, that in contrast to Figure \ref{fig:1}, Left the incoming current may consist of outgoing electrons while the outgoing current may be composed of outgoing positrons. This current brakes the rotation of the star through the torque created by the Lorentz force density $\vec j \times \vec B/c$ in the stellar atmosphere. Stellar angular momentum is then transported by the electric current (which twists the magnetic field) in the wind, and dumped somewhere far out in the wind. Where, is the big question.

The main problem with the pulsar wind is that apparently the value of $\sigma$ in (\ref{eq:1}) changes from $\sigma>>1$ near the star to $\sigma<<1$ somewhere in the wind. The first value follows from the assumption that the radio pulsar is a strongly magnetized neutron star with surface fields $B_\star \sim 10^9 - 10^{13}$ G whereas observations of the Crab nebula demonstrate that $\sigma \sim 2 \cdot 10^{-3}$ in the un-shocked wind (Figure \ref{fig:1}, Right). To solve this problem one needs to know where and how the electric current closes in the wind so as to dump the Poynting flux in the form of kinetic energy of the wind.   

Our approach to the pulsar wind problem starts with the 1D, non-relativistic, stellar wind description in terms of an electric circuit in Sect.~\ref{1D}. We then include 2D effects in Sect.~\ref{2D}, consider the aligned magnetic rotator in vacuo in Sect.~\ref{vacuum}, discuss numerical results for the electric circuit in the aligned rotator in Sect.~\ref{numerical},  consider the importance of current starvation in Sect.~\ref{starvation}, consider the multiple effects of obliquity on the electric circuits in Sect.~\ref{obliquity}, critically review the nature of current sheets in Sect.~\ref{sheets}, mention the effect of a 3D instability in Sect.~\ref{3D}, and conclude the first of the three parts of this review by summarizing our findings about the pulsar electric current circuit in Sect.~\ref{conclusion1}.

\subsection{Stellar winds: 1D MHD, partial current closure}
\label{1D}
\begin{figure}
\includegraphics[width=0.55\textwidth]{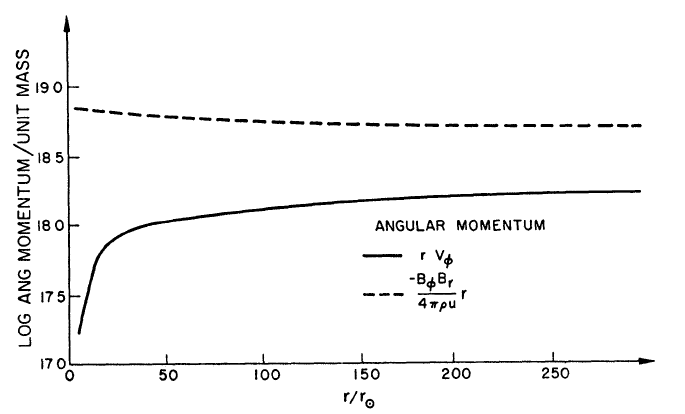}
\includegraphics[width=0.45\textwidth]{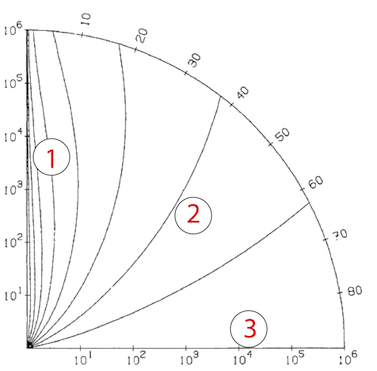}
\caption{{\bf Left}: Partition of wind specific angular momentum over kinetic and magnetic contributions with distance, expressed in solar radii, for the Sun in 1D approximation; dashed line is magnetic torque and drawn line kinetic angular momentum. The Alfv\'en radius is between 30-50 solar radii. It follows that the torque remains ultimately electromagnetic, and that only a smaller part of the current closes at a finite distance. From \protect\citet{Weber:1967fk}, reproduced by permission of the AAS. 
{\bf Right}: Focussing of a stellar wind in 2D approximation. As a result, there is subsequent complete conversion of Poynting into kinetic angular momentum. The computation is for an aligned rotator, an initially split monopole magnetic field, and an ideal MHD plasma. The bending back of the field lines towards the axis is not real and due to the logarithmic plotting of distance of a field line as a function of polar angle. Distance is in Alfv\'en radii. Credit: \protect\citet{Sakurai:1985uq}, reproduced with permission \copyright ESO.}
\label{fig:2}      
\end{figure}

Already in the early days of stellar MHD, \citet{Schatzman:1959fk} and \citet{Mestel:1968fk} realized that magnetic fields anchored in a star and extending into its atmosphere force an ionized stellar wind to co-rotate with the star to distances much larger than the photosphere. As a result the specific angular momentum carried off by a magnetized wind is much larger than its specific {\it kinetic} angular momentum at the stellar surface because of the increased lever arm provided by the magnetic field.  Our starting point is the 1D MHD description of a magnetized stellar wind by \citet{Weber:1967fk}.  Their aim is to find a steady-state smooth wind. They assume the validity of ideal (i.e. non-resistive) MHD and consider an axially symmetric magnetic structure of a special kind - a so-called split magnetic monopole - which is obtained  by reversing the magnetic field of a magnetic monopole in one half-sphere. This structure rotates around the axis of symmetry ({\it aligned} rotator), and it is assumed that the poloidal structure is not changed when rotation sets in. Finally, they consider the outflow just above the equatorial plane. They define an {\it Alfv\'en} radius $r_A$ in the equatorial plane where the ram pressure of the 
radial flow part equals the 
radial magnetic tension. More generally, the Alfv\'en radius is determined by the {\it poloidal} components \citep{Mestel:1999kx}(Ch. 7):   
\begin{equation}
\rho(r_A) v^2_{pol}(r_A)=\frac{B^2_{pol}(r_A)}{4\pi},
\label{eq:4}
\end{equation}
or equivalently, where the poloidal flow speed equals the poloidal Alfv\'en speed (i.e. an Alfv\'enic Mach number unity). They find that the angular momentum loss per unit mass equals
\begin{equation}
r v_\phi - \frac{rB_rB_\phi}{4\pi \rho v_r}=\Omega_\star r^2_A. 
\label{eq:5}
\end{equation}
Here $r, \phi$ are, formally,  spherical radius, and  azimuth  respectively, but note that their derivation is valid just above the equatorial plane ($\theta=\pi/2$) so that $r$ really is the cylindrical distance. Again the mass density $\rho$ and the other quantities are measured in the observer (lab) frame. Their main finding then is that the total torque of a magnetized wind is equal to producing effective co-rotation of the wind out to the Alfv\'en radius. Nice and simple though this result is, it is also deceptive in that the current does not close at the Alfv\'en surface. Actually, most of the current goes off to infinity. The authors find that the distribution of angular momentum over electromagnetic and kinetic parts with distance follows Figure \ref{fig:2}, Left which shows that only a part of the%
\begin{figure}
\includegraphics[width=0.55\textwidth]{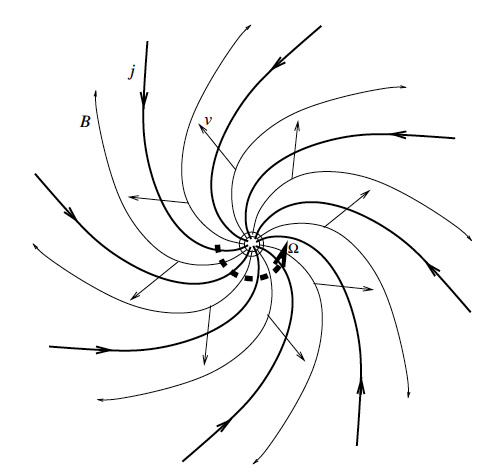}
\includegraphics[width=0.45\textwidth]{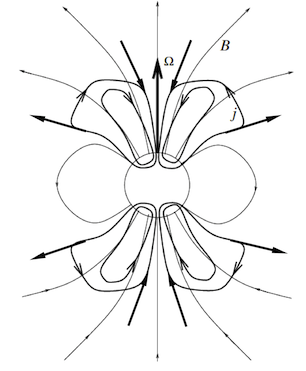}
\caption{{\bf Left}: 2D sketch of the magnetic wind of an axially symmetric, aligned rotator just above the equatorial plane as seen by a poleward observer. $B$ is the magnetic field, $v$ the wind speed, $j$ the electric current density, and $\Omega$ the rotation rate. {\bf Right}: Sketch of the same structure as projected onto the meridional plane. Note that part - but not all - of the current is closing and thereby accelerating the wind close to the star. Adapted from \protect\citet{Kuijpers:2001jl}, \copyright Cambridge University Press, reprinted with permission.}
\label{fig:3}      
\end{figure}
Poynting torque is converted into a kinetic torque, and that asymptotically at large distance the Poynting contribution becomes constant and dominates. For the current closure in the wind then (which necessarily requires more than one dimension to describe)  we can conclude that there is only {\it partial current closure}, and that most of the current goes out into the wind along the field lines in a force-free manner, i.e. asymptotically the remaining current flows along the magnetic field lines (Figure \ref{fig:3}, Right). Indeed, \citet{Weber:1967fk} find that the current within a magnetic surface satisfies
\begin{equation}
I(r)= \frac{c}{2} B_rr^2\frac{v_\phi/r-\Omega_\star}{v_r},
\label{eq:6}
\end{equation}
and becomes constant at large distances. Therefore, the result of this study - which turns out to be applicable to the wind from a {\it `slow magnetic rotator', here defined as emitting a wind mainly driven by thermal gas pressure} - is that at large distances the magnetic angular momentum and the Poynting energy flux dominate, so that $\sigma_\infty > 1$. In the next sections, we will investigate how the introduction of more realistic conditions and the transition to the pulsar wind open up the possibility of a small magnetization at infinity. 

\subsection{Stellar winds: 2D, complete current closure}
\label{2D}
The next major step in our understanding of a magnetized wind was taken by \citet{Sakurai:1985uq} who investigated the angular dependence of a steady, axially symmetric, stellar wind of an aligned rotator, again under the assumption of smooth, ideal MHD, and for an initially split monopole.  Now, it is not sufficient to solve the 1D equation of motion (the {\it Bernoulli equation}) but at the same time the equation describing force balance in the meridional plane (the {\it trans-field equation}) is required. An important result of his study is that the entire Poynting flux is converted into kinetic energy. This result can be understood from the requirement of force-balance across the poloidal field. Figure \ref{fig:3}, Right demonstrates the focussing of a magnetic wind towards the rotation axis. Three regimes can be distinguished in the wind solution: 
\begin{enumerate}
\item In the polar region, a dense jet exists where gas pressure $-\nabla p$ is balanced by the contributions from the Lorentz force $-\nabla\frac{B^2_\phi}{8\pi}$ and $\frac{(\vec B_\phi \cdot \nabla) \vec B_\phi}{4\pi} $; 
\item At intermediate latitudes, the magnetic field is force-free $\vec j \times \vec B=0$, i.e. the magnetic pressure $-\nabla\frac{B^2_\phi}{8\pi}$ and the hoop stress $\frac{(\vec B_\phi \cdot \nabla) \vec B_\phi}{4\pi} $ are in balance; 
\item Finally, in equatorial regions, the magnetic pressure $-\nabla\frac{B^2_\phi}{8\pi}$ balances the inertial force $-\rho (\vec v \cdot \nabla) \vec v$. 
\end{enumerate}
The result of magnetic focussing away from the equatorial plane is an inertial current density, given by the MHD momentum equation:
\begin{equation}
\vec j=c\frac{\vec f \times \vec B}{B^2},
\end{equation}
driven by the inertial force density 
\begin{equation}
\vec f = -\rho (\vec v \cdot \nabla) \vec v.
\end{equation} 
Since the magnetic field has both a poloidal and a toroidal component the Lorentz force associated with the inertial current density not only focusses the wind in the poloidal plane toward the rotation axis but also accelerates the stellar wind radially outward, thereby converting magnetic into kinetic energy.

\paragraph{MHD integrals for axially symmetric cold winds.}
The axially symmetric case of an MHD plasma is especially illuminating because of the four integrals of motion allowed by the (2D) MHD equations \citep{Mestel:1968fk}. Here we consider cold gas, neglect gravity, but allow for relativistic motion (Lorentz factor $\Gamma$, axial distance $r$). In terms of quantities in the laboratory frame, these conserved quantities can be written as \citep{Mestel:1999kx}(Ch. 7): 
\begin{equation}
\frac{B_\phi}{B_{pol}}=\frac{v_\phi-\Omega(\psi)r}{v_{pol}}
\label{eq:6a} 
\end{equation}
This is the {\it frozen-in field condition} which derives from the requirement that the gas exerts no stresses on the magnetic field. The parameter $\psi$ labels the magnetic flux surfaces. Expressed in the co-rotating frame in which the field pattern is static, it amounts to requiring the gas to be moving along magnetic field lines.
\begin{equation}
\frac{B_{pol}}{4\pi\rho v_{pol}}=cF(\psi)
\label{eq:6b} 
\end{equation}
This equation is tantamount to a {\it constant mass flux} along a unit outgoing poloidal flux tube.
\begin{equation}
r\Gamma v_\phi-cF(\psi)rB_\phi=L(\psi)
\label{eq:6c} 
\end{equation}
If one multiplies this equation with the constant mass flux per unit poloidal flux tube one obtains the {\it conservation of total angular momentum flux} per unit poloidal flux tube as the matter moves out. This angular momentum is made up out of kinetic specific angular momentum and electromagnetic angular momentum.
\begin{equation}
\Gamma c^2 -c F(\psi)r \Omega(\psi)B_\phi=c^2W(\psi)
\label{eq:6d} 
\end{equation}
Similarly, from this equation follows a generalization of Bernoulli's equation, the {\it conservation of total energy flux} per unit poloidal flux tube (for the cold gas) as the gas moves out. This total energy flux resides in kinetic (mass) energy flow and in Poynting flux. 

\subsection{Pulsar: aligned rotator; vacuum versus plasma}
\label{vacuum}
Before we are in a position to study the degree of magnetic focussing in a pulsar wind we have to explain why an {\it aligned} magnetic rotator is relevant at all to the radio pulsar wind. Of course, a radio pulsar only exists if the magnetic rotator is {\it oblique}, or at least if deviations from axial symmetry exist. However, there are separate important electromagnetic effects  which come in already for an aligned relativistic wind apart from the effects of obliquity, and we will try to disentangle these. Two basic consequences from Maxwell's theory require our attention:
\begin{itemize}
\item In vacuo, an axially symmetric magnetic rotator is - in a steady state -  surrounded by an axially symmetric magnetic field.  Since the stellar environs are assumed to be a vacuum no electric current flows.  Any external electric fields are axially symmetric and therefore time-independent. Therefore, there is no displacement current. The absence of both material and displacement electric currents imply that the surrounding magnetic field is purely poloidal. Since the magnetic field is time-independent any electric field has to be poloidal as well. A {\it circulating} Poynting flux does exist in the toroidal direction but there is {\it no outgoing} Poynting flux. A fortiori then, there are no radiative electromagnetic losses. 
\item In the presence of an ionized wind, the situation is different. Now, poloidal electric currents do exist, both as a result of the unipolar induction by the rotating magnet, and because of the drag on the wind from the rotating field. As a result, the external magnetic field has a toroidal component. Also, because of the ideal MHD condition (\ref{eq:2a}), an electric field appears in the lab frame which has a poloidal component. These electric and magnetic field components imply an outgoing Poynting flux (as in a common stellar wind). The magnetic field and the gas motion are, however, again time-independent, and as a result the electric field is time-independent as well. Therefore, there is no displacement current, and again no radiative losses (such as will be the case for an oblique rotator). 
\end{itemize}
Thus, by confining ourselves to the aligned rotator first, we postpone the discussion about the importance of displacement electric fields, and isolate the Poynting flux which appears in the ideal MHD approximation. Such a pulsar wind is the (relativistic) extension of the above magnetized stellar wind. 
The electric circuit is expected to close along the path of least resistance, which means across the magnetic field inside the star and along the magnetic field in the wind. However, an important difference with an ordinary star appears. The neutron star may not be able to provide sufficient gas to short out the electric field component in the wind along the magnetic field. This happens, when the local charge density is not everywhere equal to the GJ charge density~(\ref{eq:2}). There and then parallel potential drops develop.

\subsection{Aligned rotator: numerical results on collimation and acceleration}
\label{numerical}
\begin{figure}
\includegraphics[width=0.4\textwidth]{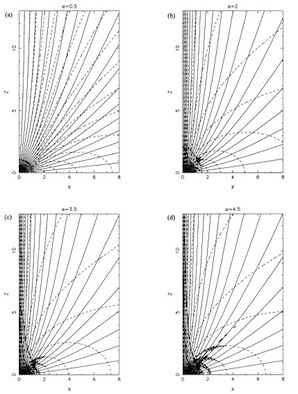}
\includegraphics[width=0.6\textwidth]{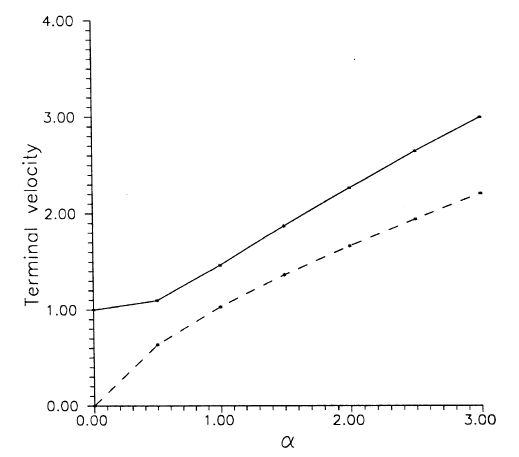}
\caption{{\bf Left}: Sequence of shapes of the poloidal field lines (drawn) with  increasing magnetic rotator parameter $\alpha$ from $\alpha = 0.5$ (solar-wind-type slow magnetic rotator) to
$\alpha= 4.5$ (fast magnetic rotator). The initial non-rotating monopole magnetic field has a spherical Alfv\'en surface located at one Alfv\'en radius ($r_A$). Distances are in
units of $r_A$ with the base located at $x = 0.5$. Dotted lines indicate poloidal currents. Thick lines indicate Alfv\'en and fast critical surfaces. From \protect\citet{Bogovalov:1999kx}, Figure 2. {\bf Right}: The terminal velocity as a function of $\alpha$ (solid line). 
For comparison, the corresponding terminal speed in Michel's minimum energy solution is also plotted (dashed line). 
From \protect\citet{Bogovalov:1999kx}, Figure 5.}
\label{fig:4}      
\end{figure}

\begin{figure}
\includegraphics[width=0.4\textwidth]{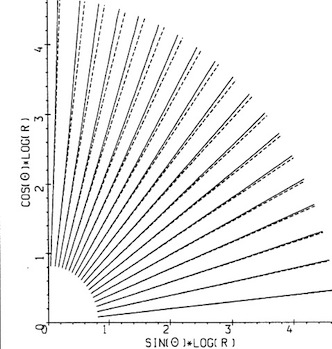}
\includegraphics[width=0.6\textwidth]{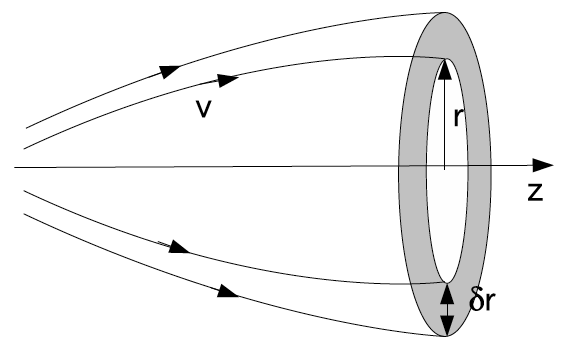}
\caption{{\bf Left}: Inefficient collimation and conversion of Poynting angular momentum in the far zone of a rotating magnetic rotator ejecting relativistic plasma. The shape of the poloidal magnetic field lines is given by solid lines. For comparison are drawn lines
of pure radial outflow (dashed). From \protect\citet{Bogovalov:1999kx}, Figure 13.
{\bf Right}: Continuous relativistic acceleration requires differential collimation which increases towards the axis.  From \protect\citet{Komissarov:2011fk}, \copyright SAIt, reproduced by permission.}
\label{fig:5}      
\end{figure}

Does the relativistic nature of the pulsar wind promote collimation and reduction of $\sigma$? The effects of fast rotation on collimation and acceleration of both  non-relativistic and relativistic stellar winds have been investigated numerically by \citet{Bogovalov:1999kx}. They find both the collimation and the  acceleration of the wind to increase with the {\it magnetic rotator parameter} \citep{Bogovalov:1999vn}
\begin{equation}
\alpha\equiv\frac{\Omega r_A}{\Gamma_0 v_0}.
\end{equation}
Here, $\Omega$ is the wind angular rotation rate (at most equal to the stellar rotation rate), $v_0$ the outflow speed at the photosphere, $\Gamma_0$ is the corresponding Lorentz factor, and $r_A$ the Alfv\'en radius. The behavior of the non-relativistic outflow is shown in Figure \ref{fig:4}, Left. Since  
relativistic outflows such as in pulsars effectively have $\Gamma_0 \sim 100$ they belong to the domain of slow rotators with $\alpha \ll 1$, and collimation becomes ineffective (Figure \ref{fig:4}, Right; Figure \ref{fig:5}, Left).

The precise role of collimation for relativistic wind acceleration has been elucidated by \citet{Komissarov:2011fk}. Consistent with his numerical results, he shows that {\it differential} collimation as sketched in Figure \ref{fig:5}, Right, is required to obtain high Lorentz factors. Such differential collimation is associated with increasing conversion efficiency of Poynting into kinetic energy flux. Tentatively, we conclude that aligned relativistic rotators lack sufficient differential collimation to establish substantial conversion of Poynting flux if the ideal MHD limit - including inertia - applies throughout the wind. Clearly, the relativistic wind motion does not help to reduce the wind magnetization. 

\begin{figure}
\includegraphics[width=0.5\textwidth]{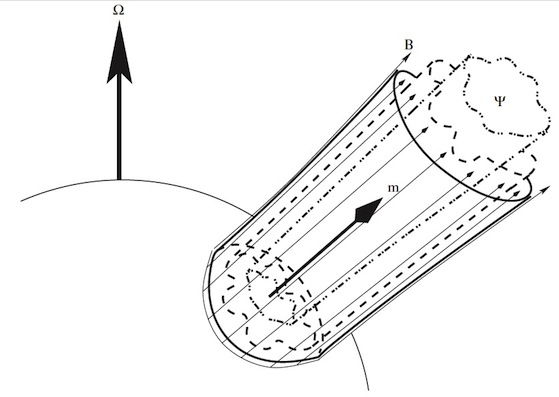}
\includegraphics[width=0.5\textwidth]{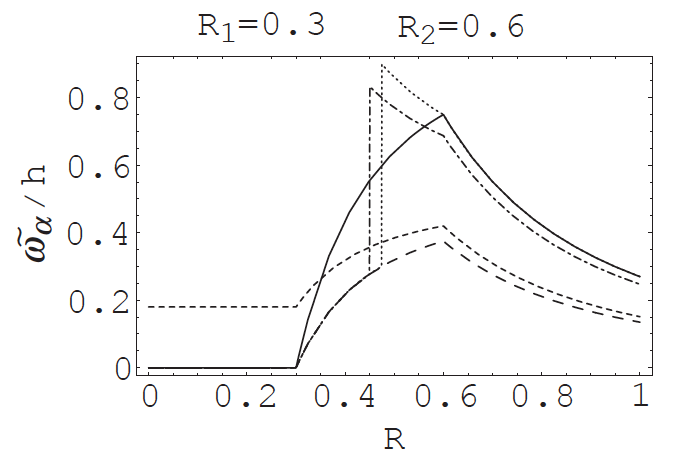}
\caption{{\bf Left}: In an oblique magnetic rotator a potential gap at the base of the open field lines does not lead to an average rotation of plasma different from stellar rotation but causes differential rotation over flux surfaces around the magnetic axis of the polar cap. Credit: \protect\citet{Fung:2006qf}, reproduced with permission  \copyright ESO.
{\bf Right}: 
Examples of differential rotation in charged, outflowing beams of mixed electrons and positrons around the magnetic axis ($z$-direction) above the polar cap of an aligned magnetic rotator and depending on the electric current density and charge density distributions. Computed are the equilibrium angular velocities {\it with respect to the lab frame} as a function of axial distance $R$ for {\it hollow} beams of mixed electrons and positrons, flowing out at approximately the same (relativistic) speed, with or without {\it core}, and with or without return current, for cylindrically symmetric beams and a uniform background magnetic field. Dimensionless angular velocity is plotted as a function of dimensionless axial distance. The core region extends from the rotation axis to $R_1=0.3$, the hollow beam(s) from $R_1=0.3$ to $R_2=0.6$, and the pulsar magnetosphere (which co-rotates with the star) starts at $R_3=1$.  Dimensionless angular velocity is given by $\tilde{\omega}_\alpha (R) \equiv \omega_\alpha (R) / \Omega_\star$. Dimensionless positron/electron charge excess in the hollow beam(s) is given by $h\equiv (n_+-n_-)/n_{GJ}$ where the GJ density is given by $n_{GJ}\equiv \tau_{GJ}/|e|$ from (\ref{eq:2c}). Defining the parameter $Q\equiv (-\beta^0_z j_z)/(j_{GJ})$, where $\beta^0_z\equiv v_z/c$ and $j_{GJ}$ is defined in (\ref{eq:2}), the various curves are for: {\bf solid}: a hollow beam with $Q = 0$; {\bf long dash}: a hollow beam with $Q = 0.5$; {\bf short dash}: a beam with $Q=0.5$ and with a core $h_I = 0.2, Q_I = 0.1$; {\bf dotted}: a beam with $Q=0.5$ surrounded by a neutralizing return current; {\bf dash-dotted}: a beam with $Q=0.5$ and a surrounding return current of the same radial extent. 
Credit: \protect\citet{Fung:2006qf}, reproduced with permission  \copyright ESO. }
\label{fig:6a}      
\end{figure}

\paragraph{The effect of a parallel potential gap.}
\citet{Contopoulos:2005zr} shows that a constant potential gap at the basis of the open field lines allows the (otherwise ideal) wind to sub-rotate with respect to the stellar rotation, assuming an {\it aligned}, initially split monopole  
\begin{figure}
\includegraphics[width=0.55\textwidth]{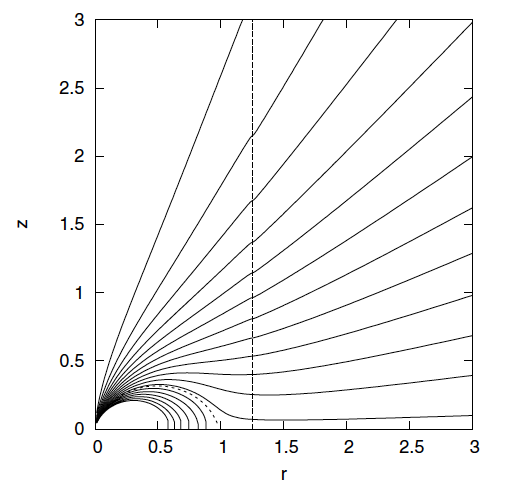}
\includegraphics[width=0.45\textwidth]{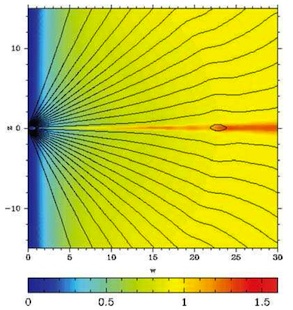}
\caption{{\bf Left}: Poloidal field lines (drawn) for an aligned rotator with a potential gap at the base, and for otherwise ideal MHD. The wind relative rotation rate is 0.8. The poloidal flux increases between field lines in steps of 0.1, counting from zero on the vertical axis. The dotted line shows the separatrix (boundary between open and closed field lines) at a flux value of 1.23. Distance is expressed in light cylinder radii. Credit: \protect\citet{Contopoulos:2005zr}, reproduced with permission  \copyright ESO.
 {\bf Right}: The wind zone structure of a dipolar magnetosphere at much larger distance than in the Figure on the left is again practically radial. The contours show the field lines of poloidal magnetic field. The color image shows the distribution of the logarithm of the Lorentz factor. From \protect\citet{Komissarov:2006ly}, Figure 7.}
\label{fig:6}      
\end{figure}
for the poloidal field (Figure \ref{fig:6}, Left). However, two remarks are in place.  
First, both \citet{Contopoulos:2005zr}  and \citet{Timokhin:2007oq} neglect the influence of the outgoing/ incoming current on the drift speed. In case of a current, the general expression for the drift speed with respect to the lab frame is given by (cylindrical coordinates $z, R, \phi$) \citep{Fung:2006qf}\begin{equation}
v_{drift}= -c\frac{E_R}{B_0}+v_z\frac{B_\phi}{B_0},
\label{eq:12a}
\end{equation}  
where $B_\phi$ is generated by the current itself. For a relativistic plasma as in the pulsar wind, the final term cannot be neglected, and, in fact can lead to a near-cancellation of the drift so as to cause the gas not to co-rotate with the pulsar at all (Figure \ref{fig:6a}, Right).
Secondly, as can be seen in \citet{Fung:2006qf} differential rotation of the open field lines takes place around the magnetic axis, not  around the rotation axis. Indeed, in an oblique rotator, a potential gap leads to circulation of the open field lines 
around the oblique magnetic axis, superposed on a pattern which still is rotationally locked to the star (Figure \ref{fig:6a}, Left). It is interesting to note that in the terrestrial magnetosphere (where rotation is unimportant) differential rotation is also oriented around the dipole axis, not around the rotation axis.

\paragraph{More numerical MHD results.}
\citet{Komissarov:2006ly} models the aligned split monopole with relativistic ideal MHD, including inertia, artificial resistivity, and artificially resetting gas pressures and densities. His spatial domain extends out to a much larger distance than in \citet{Contopoulos:2005zr} (Figure \ref{fig:6}, Right). Again, his computations do not show any collimation of the field lines towards the rotation axis. Neither do they solve the issue of current closure nor the conversion of Poynting flux as required by the observations of $\sigma$. 

\subsection{Current starvation and Generalized Magnetic Reconnection}
\label{starvation}
\begin{figure}
\includegraphics[width=0.8\textwidth]{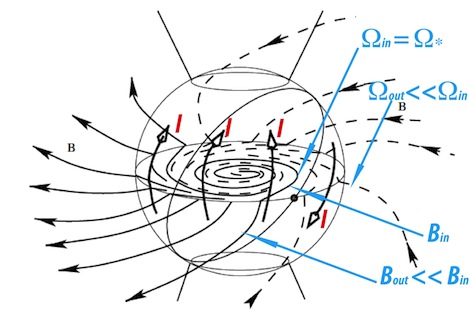}
\caption{Sketch of the magnetic field lines (black arrows) and of the electric current lines (open arrows) far out in the wind where current starvation occurs. Here strong electric fields are generated parallel to the ambient magnetic fields, the electric current closes and dissipates, thereby converting most of the Poynting angular momentum into kinetic, and the wind is strongly heated and accelerated.  As a result, the toroidal magnetic field component drops sharply. Outside the dissipative layer the field pattern rotates much slower and slips over the inside field structure. From \protect\citet{Kuijpers:2001jl}, \copyright Cambridge University Press, reprinted with permission.}
\label{fig:7}      
\end{figure}

The solution to the problem of current closure in an aligned rotator may be given by what is called current starvation. If one insists on a force-free wind the total electric current is conserved if one follows the circuit outward since the current density is everywhere parallel to the open magnetic field lines. Since these field lines become increasingly toroidal in the outward direction the same happens with the electric current density. A conserved current $I$ then implies the following dependence of current density with distance $r$ to the pulsar:
\begin{equation}
j \approx \frac{I_{GJ}}{ r \pi r_{LC}\delta},
\label{eq:13a}
\end{equation}
where $\delta$ is the asymptotic opening angle between two selected magnetic flux surfaces.
However, the
available current density falls off as (quantities in the co-moving frame are dashed) 
\begin{equation}
j_{max}= j'_{max}=n' e c=\frac{nec}{\Gamma} \propto \frac{1}{r^2}.
\label{eq:13b}
\end{equation}
Here we have assumed that the (force-free) current density is mainly toroidal (as is the magnetic field), and therefore invariant under a Lorentz transformation to the co-moving frame, while the density is not and obtains a Lorentz factor upon transformation back to the observer frame. Apparently, since the drift speed runs up against the speed of light  a current starvation problem arises    
at a radius well within the termination shock \citep{Kuijpers:2001jl}
\begin{equation}
\frac{R_{max}}{r_{LC}}\approx\frac{M}{2\Gamma}.
\end{equation}
Here $r_{LC}\equiv c/\Omega_\star$ is the light cylinder radius, $\Gamma$ is the wind Lorentz factor (before current closure), and $M\equiv n/n_{GJ}$ is the multiplicity of the pair plasma in the wind expressed in terms of the GJ density at the base \\
$n_{GJ}\equiv \vec \Omega_\star \cdot \vec B (2 \pi e c)^{-1}$. 
Such a shortage of charges leads 
to strong electric fields parallel to the ambient magnetic
fields which try to maintain the current. The effect of the parallel electric fields can be described as {\it Generalized Magnetic Reconnection}, a term introduced by
\citet{Schindler:1988kx} and \citet{Hesse:1988uq}. In this region ideal MHD must break down and particles are accelerated (heated) along the magnetic field. At the same time, the parallel electric field causes the external magnetic field pattern to slip over the inner fast rotating pattern at a much smaller rate. This implies a much smaller external toroidal magnetic field, and this implies a much smaller current. In other words, the current closes across the field, and the associated Lorentz force accelerates the wind in both radial and toroidal directions. As a net result the current dissipates in this layer and causes both heating and bulk acceleration of the wind.  A sketch of such a dissipative shell is given in Figure \ref{fig:7}. This current closure due to current starvation is a natural candidate for  
the main conversion of the Poynting flux in the MHD wind of a pulsar into kinetic energy  \citep{Kuijpers:2001jl}. Finally, note that this current starvation occurs throughout the wind and already for the aligned pulsar, which is different from the current starvation in \citet{Melatos:1997fk} who considers the singular current layer which he postulates to occur at the site of the displacement current, and which exists for an oblique rotator only.

\subsection{Oblique rotator: what to expect?}
\label{obliquity}
\begin{figure}
\includegraphics[width=0.8\textwidth]{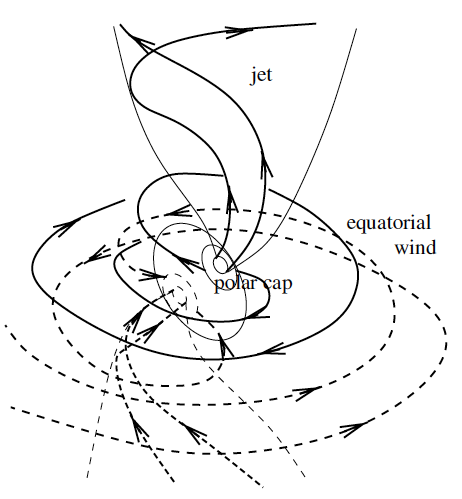}
\caption{Sketch of a relativistic oblique magnetic rotator. Three magnetic domains can be distinguished: two polar jets, each with magnetic field of one polarity, and an equatorial wind with alternating flux bundles originating at the respective magnetic poles. From \protect\citet{Kuijpers:2001jl}, \copyright Cambridge University Press, reprinted with permission.}
\label{fig:8}      
\end{figure}

From the study of the aligned magnetic rotator we come to the conclusion of the applicability of 
\begin{itemize}
\item ideal and force-free MHD in the main part of the pulsar wind, \item ideal, non-force-free MHD inside the neutron star,   
\item effectively resistive MHD in the accelerating gaps above the magnetic poles,  
\item and a far-out current starvation region of current dissipation. 
\end{itemize} 
Since the formation of an MHD wind does not require a magnetic rotator to be aligned, the above considerations remain applicable to the oblique rotator, be it in a modified form. Further, in an oblique rotator also new effects appear because of the change of magnetic geometry. In particular the time-dependence causes the appearance of a displacement current which is absent in the aligned case. The overall magnetic geometry is sketched in Figure \ref{fig:8}. 
Although axial symmetry is lost on a small scale, still the rotation together with magnetic (hoop) stresses are expected to lead to some global axial symmetry in two polar jets each of one (opposite) polarity only. In between, there is an equatorial wind consisting of alternating `stripes' of magnetic field combining both polarities \citep{Coroniti:1990zr}. All three domains contain electric currents. However, whereas the magnetic field inside a jet varies smoothly, the magnetic field in the equatorial wind has an alternating `striped' spiral structure, and is therefore strongly time-dependent. As a result, a new phenomenon appears in the equatorial wind, the displacement current  which usually is absent in a non-relativistic MHD approximation.

\paragraph{Displacement current in an oblique rotator wind.}
\citet{Melatos:1996ve} point out that, in an oblique rotator, the displacement current density $\partial \vec E /(c\partial t)$ increases with respect to the `conduction' current density $\vec j$ with distance from the star:
\begin{equation}
\left|\frac{\frac{\partial \vec E}{c\partial t}}{\vec j}\right|>\frac{eE\Omega_\star}{mc\omega_0^2} \{1, \Gamma\} \propto r,
\end{equation}
where it is assumed that $E\propto 1/r$ and the total plasma frequency is given by
\begin{equation}
\omega_0^2 \equiv \frac{4\pi (n^{+}+n^{-})e^2}{m_e}\propto \frac{1}{r^2}.
\end{equation}
Here the first term inside curly brackets applies to a current transverse to the relativistic outflow and the second term to a parallel current. For the Crab Nebula with 
$r_{shock}=2\times 10^9\; r_{LC}$ the displacement current density becomes larger than the charged current density at an estimated critical radius of
$r_{cr}=1\times 10^5\; r_{LC}$ in case of a transverse current. We note that this critical radius is larger than our estimate for current starvation for Crab values $M\sim 10^5$ and $\Gamma \sim 10^2$ so that current starvation remains important. Their point, however, that the displacement current cannot be neglected is well made.

\paragraph{Force-free wind of an oblique rotator contains a displacement current sheet.}
Appealing though the concept of a striped wind \citep{Coroniti:1990zr, Bogovalov:1999vn} is, an inconsistency exists in its description since the assumption of the ideal MHD condition (\ref{eq:2a}) to be valid everywhere leads to a contradiction. The wind speed is continuous over each stripe while the magnetic field reverses sign (Figure \ref{fig:8}), and therefore the electric field as given by (\ref{eq:2a}) is a step-function in time for a static observer. It then follows immediately from Amp\`ere's law that there is a singular sheet of displacement current. In a realistic situation, the width would be finite and represents a strong radiative pulse which is not discussed at all. Instead, the striped wind concept studies the energy liberated in this sheet by reconnection.

In a follow-up \citet{Bogovalov:1999vn} presents the transition between two stripes as a tangential discontinuity with a charge current (Figure \ref{fig:17}) instead of a displacement current. He constructs this equatorial current sheet by picking out field lines starting in the equatorial magnetic plane from 
\begin{figure}
\includegraphics[width=0.55\textwidth]{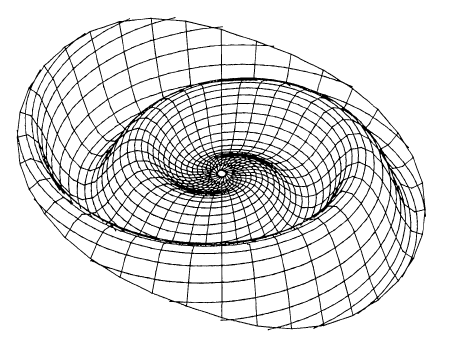}
\includegraphics[width=0.45\textwidth]{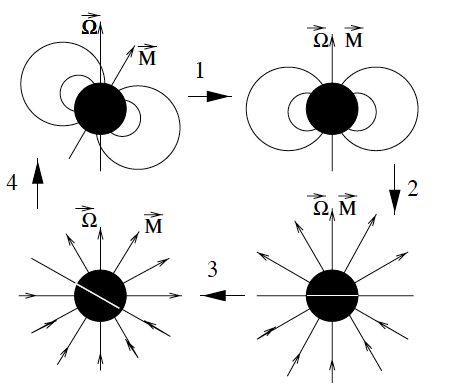}
\caption{{\bf Left}: Sketch of the `ballerina' curtain separating the polarities of the magnetic field lines of a rotating oblique split monopole. The field lines constituting this separatrix surface are selected from the 2D wind model by \protect\citet{Sakurai:1985uq} for the magnetic monopole-like basic configuration by inserting a plane tilted at $10^o$ with respect to the (rotational) equator near the star and following the individual field lines starting at the cross-section of plane and stellar surface.  The magnetic field lines below this surface are then reversed. Credit: \protect\citet{Sakurai:1985uq}, reproduced with permission  \copyright ESO.  {\bf Right}: The same procedure has been used by \protect\citet{Bogovalov:1999vn} to obtain the model wind for an (initial) split-monopole: the author first considers the aligned case, then computes the wind from a corresponding magnetic monopole, and finally reverses the field line directions within a tilted stellar hemisphere. Essentially, the procedure is allowed since a monopole has no (magnetic) equator. See also Figure \ref{fig:16}, Left. Credit: \protect\citet{Bogovalov:1999vn}, reproduced with permission  \copyright ESO.}
\label{fig:9}      
\end{figure}
an initial oblique split-monopole (Figure \ref{fig:9}). 

\paragraph{Simple proof for the existence of a displacement current.}
Here we would like to point out a simple proof for the relative importance of a displacement current versus a charge current in the wind of an oblique rotator. Let us start by assuming that the electric field is determined by the ideal MHD condition everywhere (\ref{eq:2a}). Now, substitute this electric field into Amp\`ere's law.  For a static observer $|\partial \vec B/\partial t|\sim Bv/(\pi r_{LC})$ and $|\nabla\times\vec B|\sim B/(\pi r_{LC})$. One then immediately finds, contrary to the initial assumption, that the time-varying part of the wind is associated with a displacement current of relative importance (for the extreme case of a perpendicular rotator)
\begin{equation}
\frac{|\frac{1}{4\pi}\frac{\partial \vec E}{\partial t}|}{\left|\frac{c}{4\pi}\nabla\times\vec B\right|}\sim 1-\frac{1}{\Gamma^2}.
\label{eq:j22}
\end{equation}
 
Further this result implies that the stationary part of the wind - which satisfies the ideal MHD condition, and which may be subject to charge starvation, satisfies
\begin{equation}
\frac{|\vec j|}{\left|\frac{c}{4\pi}\nabla\times\vec B\right|}\sim\frac{1}{\Gamma^2}.
\label{eq:j23}
\end{equation}
Clearly, since the pulsar wind is believed to have $\Gamma \sim 100-200$ already at its base, a displacement current cannot be neglected, and in fact dominates over the charge current in the equatorial wind. This turns out to be important since a displacement current has a different way of dissipating (i.e. `Co-moving Poynting Flux Acceleration') than the charge current.

\begin{figure}
\includegraphics[width=0.47\textwidth]{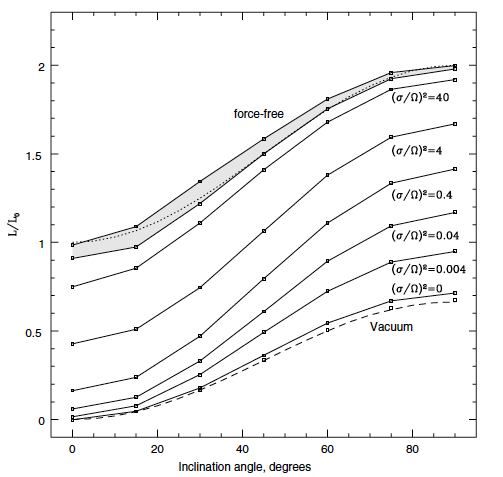}
\includegraphics[width=0.53\textwidth]{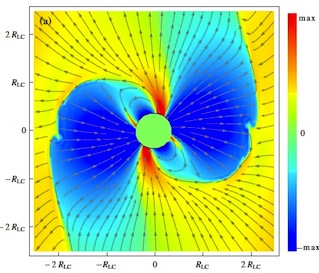}
\caption{{\bf Left}: Dimensionless energy losses as a function of inclination angle of a rotating magnetic dipolar star, and as a function of (uniform) resistivity in a resistive MHD computation. Conduction and displacement currents weaken with decreasing conductivity. From \protect\citet{Li:2012qf}, reproduced by permission of the AAS.
{\bf Right}: Force-free magnetic field lines in the plane containing both the rotation and the magnetic axes for a 60$^o$ inclined dipole. Color represents out-of-plane magnetic field into (red) and out of (blue) the page. The color table shows only values up to 30\% of the maximum of the out-of-plane magnetic field, and a square root stretching has been applied to its magnitude. Ideal force-free case. From \protect\citet{Li:2012qf}, reproduced by permission of the AAS.}
\label{fig:10}      
\end{figure}

\paragraph{Vacuum versus plasma: numerical results.}
\citet{Li:2012qf} model the oblique pulsar wind for various inclinations while neglecting inertia and assuming a uniform conductivity. The power in their wind solutions varies depending on conductivity, which ranges from infinity (the force-free case) to zero (the vacuum), and is given in Figure \ref{fig:10}, Left.  Of course, these results do not represent a real pulsar 
wind, mainly since the actual conductivity is highly varying in space and time. Although an infinite conductivity is probably a good approximation in most of space the nature of the wind depends critically on many small, self-consistently determined, non-ideal domains, e.g. the conductivity of vacuum in polar and outer wind gaps, and the anomalous resistivity in regions of pair creation and radio emission. Nevertheless, the results are instructive (Figure \ref{fig:10}, Right), in particular as to the current circuit and the relative role of Poynting flux associated with the MHD current on one hand versus displacement current on the other hand. For the case of infinite conductivity, the (wind) power of an aligned rotator - where the displacement current is absent - is found to be only half that of the perpendicular rotator where the displacement current reaches its maximum. 

\paragraph{Wind from an oblique rotator as a TEM wave.}
Does the displacement current in the wind dissipate, and reduce $\sigma$ before the shock? \citet{Skjaeraasen:2005bh} model the pulsar wind as a large-amplitude, superluminal, nearly Transverse Electromagnetic (TEM) wave in a relativistically streaming (Lorentz factor $\gamma_d$) electron-positron plasma with dispersion relation
\begin{equation}
\left(\frac {ck}{\omega}\right)^2=1-\frac{2\omega_p^2}{\omega^2\gamma_d\sqrt{1+\hat E^2}},
\end{equation}
where the plasma frequency is given by 
\begin{equation}
\omega_p^2=\frac{4\pi n^\pm e^2}{m_e},
\end{equation}
the density of electron-positron pairs is $n^\pm$, 
the dimensionless wave electric field is given by 
\begin{equation}
\hat E= \frac{eE}{mc\omega},
\end{equation}
and $\omega$ is the wave frequency.
They then study the evolution of this TEM wave when a shock confines the TEM, and find that as the 
\begin{figure}
\includegraphics[width=0.5\textwidth]{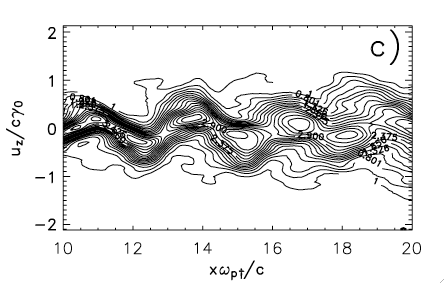}
\includegraphics[width=0.5\textwidth]{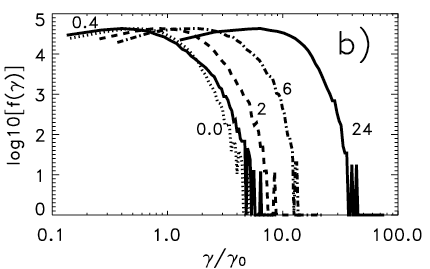}
\caption{{\bf Left}: Phase space diagram of shock precursor: a nonlinear TEM wave decays, and broadens with distance, while the thermal background spectrum grows. Initially, the particles are strongly phase-coherent
with the wave;  thermal broadening can be seen at positions beyond $x > 11$. $u$ is velocity, $\gamma$ Lorentz factor, and $\omega_p$  plasma frequency. From \protect\citet{Skjaeraasen:2005bh}, reproduced by permission of the AAS.
{\bf Right}: Energy spectra of particles with position. As the distance increases particles are accelerated to higher energies by the decaying TEM. From \protect\citet{Skjaeraasen:2005bh}, reproduced by permission of the AAS.}
\label{fig:11}      
\end{figure}
distance in the wind increases, the wave amplitude decays and its frequency width broadens, while the thermal background spectrum grows (Figure \ref{fig:11}).

\begin{figure}
\includegraphics[width=0.42\textwidth]{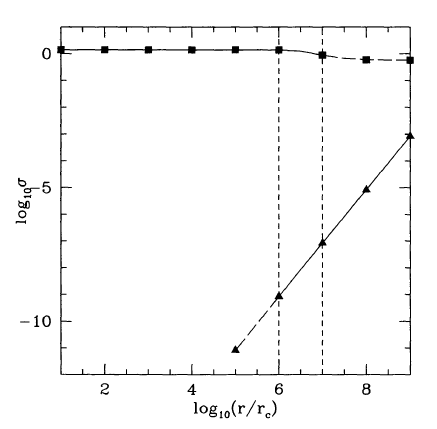}
\includegraphics[width=0.58\textwidth]{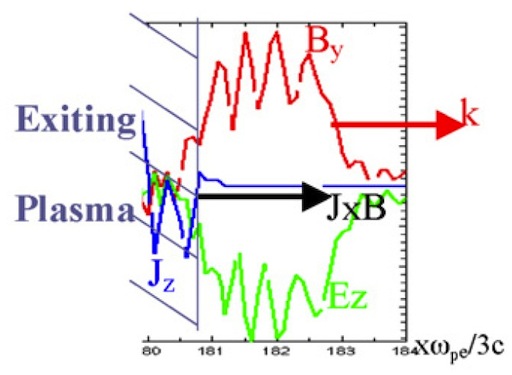}
\caption{{\bf Left}: The $\sigma$-parameter for the two exact transverse wave solutions in a pulsar wind as function of distance. Initially, the pulsar wind can be fitted to the subluminal wave (top) where the particles are magnetized in the radial magnetic field. How the transition is from a wind which is predominantly a subluminal wave to a superluminal wave (bottom), where $\sigma<<1$ and the particles are unmagnetized, is unclear. From \protect\citet{Melatos:1998uq}, \copyright SAIt, reproduced by permission.
{\bf Right}: Co-moving Poynting Flux Acceleration: A finite-width electromagnetic pulse leads to an accelerating force (3D sketch). From \protect\citet{Liang:2009cr}, reproduced by permission of the AAS.}
\label{fig:12}      
\end{figure}

\paragraph{Crab wind energy carried by a TEM wave.}
\citet{Melatos:1998uq} solves for exact transverse wave solutions and finds two circularly polarized waves, one subluminal wave in which the particles are magnetized in the radial magnetic field, and consequently do not oscillate in the wave field so that $\gamma^\pm \approx \gamma_d$, and one superluminal wave in which the particles are unmagnetized and oscillate with the wave field so that $\gamma^\pm = \gamma_d (1+\hat E^2)^{0.5} \approx \gamma_d \hat E$. Clearly, since $\hat E \sim 10^{11}$ at the light cylinder for the Crab, the superluminal wave has a small magnetization as required by the observations (see Figure \ref{fig:12}, Left). 
The problem, however, is that the properties of the pulsar wind at the base more resemble  those of the subluminal wave. How the transformation from one wave into the other takes place is not clear.

\paragraph{Co-moving Poynting Flux Acceleration (CPFA).}
What then is the main dissipation mechanism for strong electromagnetic pulses with large magnetization? Transverse EM pulses with moderate electric fields in a plasma cause particles to move at the steady drift speed $c\vec E \times \vec B/B^2$. However, for near-vacuum waves with $E\sim B$, particles can be accelerated impulsively by the ponderomotive force in the wave front, catching up with
\begin{figure}
\includegraphics[width=0.5\textwidth]{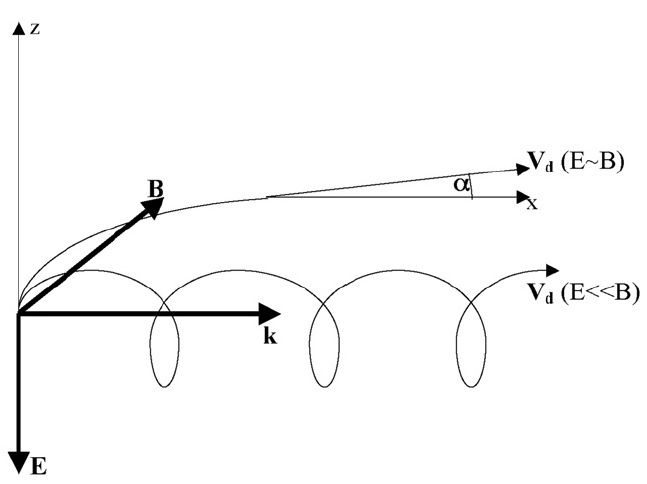}
\includegraphics[width=0.5\textwidth]{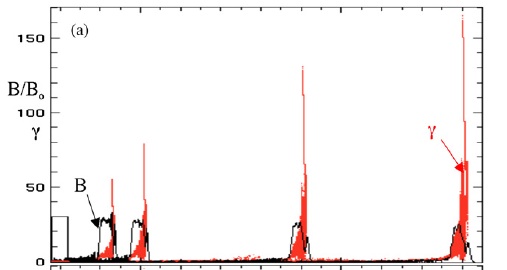}
\caption{{\bf Left}: When the electric field amplitude in an electromagnetic pulse becomes comparable to its magnetic field amplitude, the drift speed approaches the speed of light, and particles are accelerated by the ponderomotive force in the front of the pulse. This happens in the `impulsive plasma gun'. From \protect\citet{Liang:2009cr}, reproduced by permission of the AAS.
{\bf Right} Computation of impulsive particle acceleration (Lorentz factor in red) by a discrete magnetic pulse (black) as a function of time. From \protect\citet{Liang:2009cr}, reproduced by permission of the AAS.}
\label{fig:13}      
\end{figure}
and retarding the wave slightly (Figure \ref{fig:12}, Right). This process has been proposed by \citet{Contopoulos:1995nx} to operate in oblique pulsar winds (`impulsive plasma gun',  Figure \ref{fig:13}, Left).  Figure \ref{fig:13}, Right shows impulsive acceleration as computed by \citet{Liang:2009cr} for a strong EM pulse in the absence of a guiding magnetic field. Clearly, the process is effective although the Lorentz factors obtained in this study are still far below what is required in a pulsar wind.

\subsection{Current sheets.}
\label{sheets}
Within the pulsar context one often encounters the assumption of   - infinitesimally thin - current sheets. As long as these are invoked as approximate circuit elements required for current closure, there is no problem. However, when these current sheets become ingredients of dissipation it is relevant to ask how they have developed out of space-filling `body' currents, as for example one is used to do in the solar corona flare context. Below, we will discuss the nature of two main classes of current sheets invoked in pulsar studies: singular return currents at the boundary between open and closed field lines, and current sheets between the stripes in the striped wind model.

\begin{figure}
\includegraphics[width=0.35\textwidth]{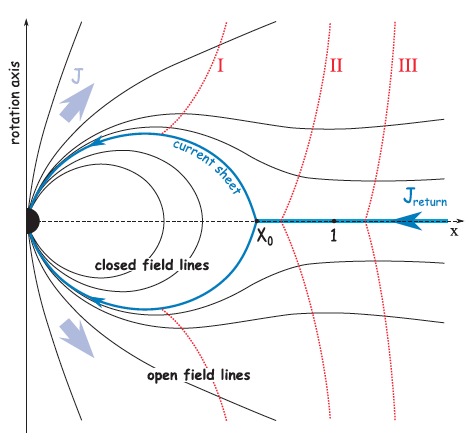}
\includegraphics[width=0.65\textwidth]{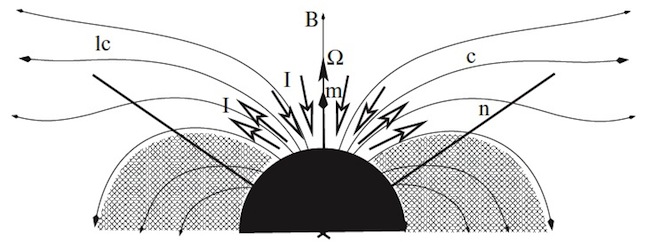}
\caption{{\bf Left}: Sheet currents in a typical theoretical model (blue). From \protect\citet{Timokhin:2007oq}, Figure 1.  {\bf Right}: Finite-width currents in original Goldreich-Julian model (polar cap is exaggerated). Credit: \protect\citet{Fung:2006qf}, reproduced with permission  \copyright ESO.}
\label{fig:14}      
\end{figure}

\paragraph{Current sheets on the boundary between open and closed field?}
Models of the magnetosphere of a magnetic rotator show distributed currents along the open magnetic field lines above the pulsar polar caps but often return currents in the form of sheets, such as indicated by arrows in Figure \ref{fig:14}, Left. The current sheet then is along the last open magnetic field line, separating zones of open 
and closed magnetic field lines. However, as we see it, this is just a simplification. By its very origin the return current is not a singular layer but a body current, just as the polar cap current. Actually, this was  discussed already in the early seminal paper by \citet{Goldreich:1969fk}, (Figure \ref{fig:14}, Right).
\begin{figure}
\includegraphics[width=0.65\textwidth]{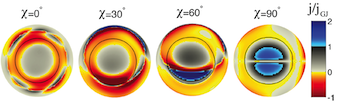}
\includegraphics[width=0.3\textwidth]{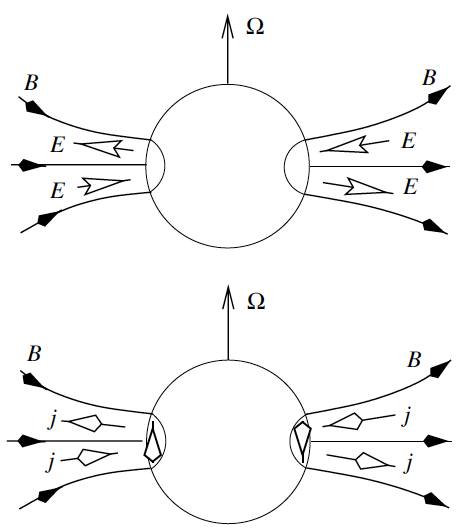}
\caption{{\bf Left}: Finite width return current in actual computation. From \protect\citet{Timokhin:2013kl}, Figure 1. 
{\bf Right}: Perpendicular rotator with electric fields (top) and electric currents (bottom). Current and return current have similar widths. From \protect\citet{Kuijpers:2001jl} \copyright Cambridge University Press, reprinted with permission.}
\label{fig:15}      
\end{figure}
Also, numerical studies find that the electric (polarization) currents to and from the pulsar in the polar cap region are quite broad as can be seen in the representative example of Figure \ref{fig:10}, Right,  which is computed for an oblique $60^o$ dipole and under the force-free assumption. Another recent numerical study \citep{Timokhin:2013kl} shows a distributed return current  with dimensions comparable to the polar cap width (Figure \ref{fig:15}, Left).  In particular, the perpendicular rotator presents an interesting example where the symmetry causes currents and return currents to have the same finite width as can be seen in Figure \ref{fig:15}, Right. Note that there is a puzzling difference between the two papers in the directions of the currents for the perpendicular rotator: in Figure \ref{fig:15}, Left, the currents in the polar cap above the equator have the same directions as those below the equator, while in Figure \ref{fig:15}, Right, the currents are opposite, and form part of one circuit. In our opinion, the latter picture is correct as it is based on the sign of the surface charges for a perpendicular vacuum rotator which is given by the sign of $\vec \Omega_{\star} \cdot \vec B$, and which is opposite on both sides of the equator.

\begin{figure}
\includegraphics[width=0.47\textwidth]{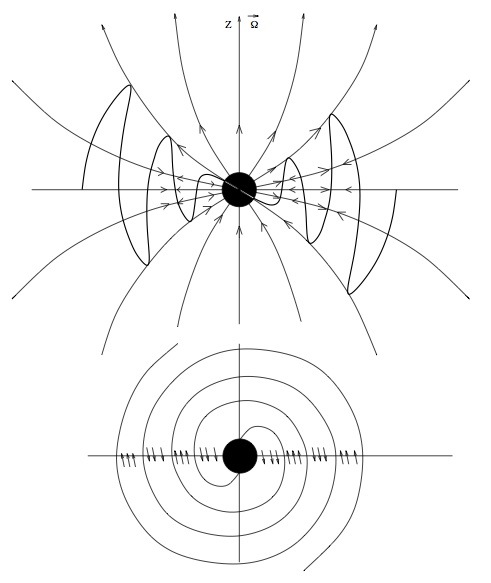}
\includegraphics[width=0.53\textwidth]{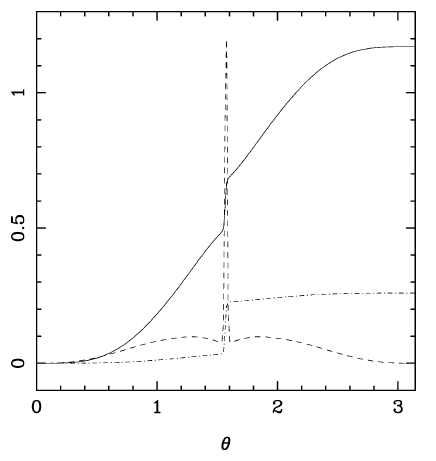}
\caption{{\bf Left}: Meridional (top) and equatorial (bottom) cross-section of the oblique split-monopole wind. Credit: \protect\citet{Bogovalov:1999vn}, reproduced with permission  \copyright ESO.
{\bf Right}: The angular distribution of the relative luminosities of a pulsar wind from an axially symmetric computation with numerical resistivity only, at $r/r_{LC}=  10$: total luminosity (solid line), hydrodynamic kinetic (dot-dashed line) within the polar cone of angle $\theta$. The
dashed line shows the total flux density in the radial direction. From \protect\citet{Komissarov:2006ly}, Figure 9 Right.}
\label{fig:16}      
\end{figure}

\paragraph{Current sheets between equatorial stripes?}
The other class of singular currents occurs at the tangential discontinuities between field bundles of different polarities when these meet in the equatorial region. The most simple case is that  of the aligned rotator (Figure \ref{fig:14}), Left, which shows the meridional projection of the equatorial tangential discontinuity. In 3D this current spirals inward in the equatorial plane. Of course, in the oblique case this flat current sheet becomes undulating (Figure \ref{fig:16}, Left). Many authors concentrate on this tangential discontinuity in the striped equatorial wind as a source of converting magnetic into kinetic energy \citep{Coroniti:1990zr,  Bogovalov:1999vn}. There are, however, two objections to be made to these studies. One is that this wrinkled current sheet is in reality made up out of a combination of the tangential discontinuity current plus the displacement current  we have discussed above. Surely, dissipation of this current sheet then cannot be handled purely from a resistive MHD point of view. The second objection is that, in confining the study to the singular layer, one neglects the `body' currents within the stripe which connect to the magnetic poles and which carry both angular momentum and energy from the star. How such body currents would condense into a singular sheet is not made clear.

\begin{figure}
\includegraphics[width=0.5\textwidth]{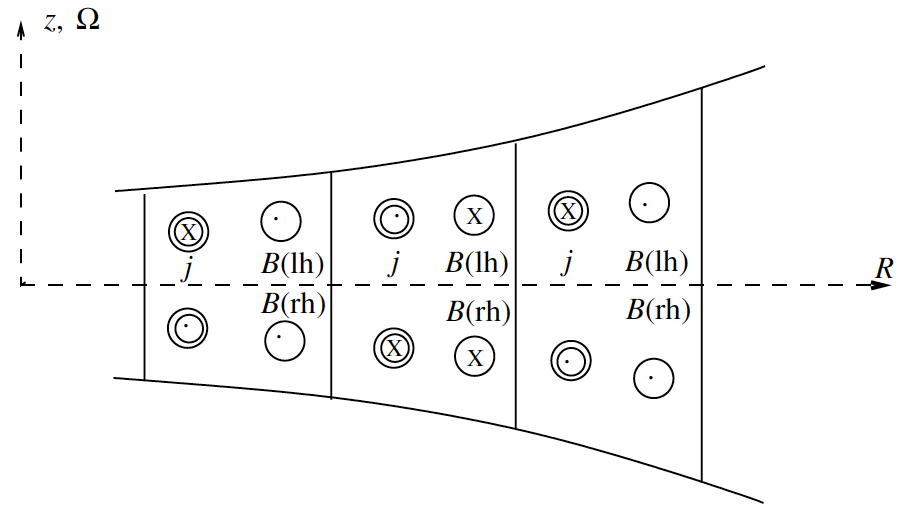}
\includegraphics[width=0.5\textwidth]{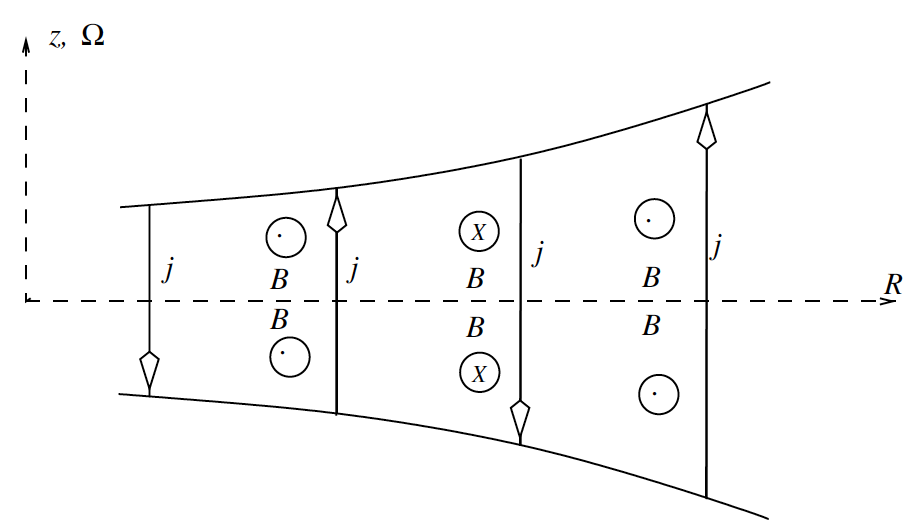}
\caption{{\bf Left}: Meridional cross-section of the equatorial wind. The field in each segment connects to the corresponding polar cap on the star. The horizontal width of a flux tube is roughly constant in the force-free cold flow, and equal to $\pi r_{LC}$. The force-free nature of the current implies that the magnetic field in each flux tube is helical as indicated in brackets (lh for left-handed; rh for right-handed). The pitch of the field on the northern hemisphere is left-handed. 
{\bf Right}: Conventional current picture which focuses
on the singular current sheets associated with the tangential  discontinuity and leaves out any body currents. From \protect\citet{Kuijpers:2001jl}, \copyright Cambridge University Press, reprinted with permission.
}
\label{fig:17}      
\end{figure}

\begin{figure}
\includegraphics[width=0.8\textwidth]{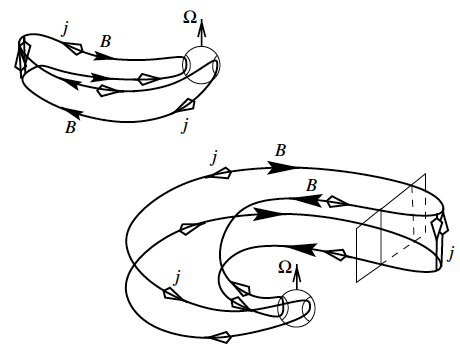}
\caption{Sketch of electric currents (open arrows) which
are set up as a flux tube connecting both magnetic poles is drawn out from a perpendicular magnetic rotator so that it brakes the stellar rotation. For clarity, only two magnetic field lines (black arrows) are shown. The inclined plane on the right is the cut presented in Figure \ref{fig:17} in the asymptotic regime when the top of the loop has vanished to infinity. From \protect\citet{Kuijpers:2001jl}, \copyright Cambridge University Press, reprinted with permission.
}
\label{fig:18}      
\end{figure}

\paragraph{Towards a realistic description of the equatorial wind of an oblique rotator.}
Figure \ref{fig:17} shows sketches of the meridional cross-section of part of the equatorial pulsar wind, according to existing lore (Right) and our view (Left).  As we see it, the equatorial region of the wind of an oblique rotator consists of mainly toroidal magnetic flux tubes which come from both poles and join up, arranged in an alternating pattern in radial direction. Each flux tube carries a body current, with opposite signs above and below the equator. These currents correspond to the internal twist of the flux tubes caused by the polar in-coming and out-going current system. We have constructed this sketch based on the evolutionary scenario of the events when a finite flux tube is drawn out from a star when it starts rotating (Figure \ref{fig:18}). Apart from these currents there is a current which separates two flux tubes. This current  consists to a negligible  amount (of order $\Gamma^{-2}$) out of a charge current, and mainly out of a displacement current. Altogether, the Poynting flux of the oblique rotator is supported by two electric current systems: a charge body current, which may be subject to current starvation and consequent dissipation at some distance from the star, and a displacement current, which probably is distributed instead of singular as well, and which dissipates by the CPA process. The third component of a postulated singular charge current separating flux tubes from both hemispheres - on which several papers focus - can be completely neglected as to its contribution to 
the total Poynting energy flux. In our opinion this is corroborated by the results from \citet{Komissarov:2006ly} who computes the amount of Poynting flux dissipated by numerical resistivity in MHD approximation 
of a pulsar wind constructed according to the recipe by \citet{Bogovalov:1999vn} (Figure \ref{fig:16}, Right). As can be seen the contribution from the equatorial current sheet is minor with respect to the total energy flux.

\subsection{3D-instability of a toroidal wind.}
\label{3D}
\begin{figure}
\includegraphics[width=0.7\textwidth]{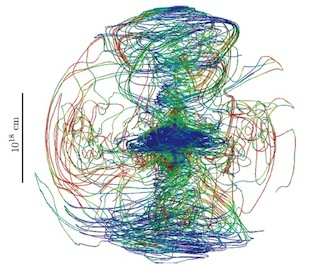}
\caption{Kink instability of toroidal field behind the shock. 3D rendering of the magnetic field structure in a model with $\sigma_0= 3$ at $t=  70$ yr after the start of the simulation. Magnetic field
lines are integrated from sample points starting at $r = 3 \cdot 10^{17}$  cm. colors indicate the dominating field component, blue for toroidal and red for poloidal. From \protect\citet{Porth:2013tg}, Figure 2 Left.}
\label{fig:19}      
\end{figure}

An intrinsically different solution to the sigma problem has been proposed by \citet{Begelman:1998uq}. He points out that the strong toroidal field amplified by the reverse shock of the wind is Rayleigh-Taylor unstable, and has the tendency to rise out of the equatorial plane. The shocked magnetic field and the electric current system are then expected to acquire a turbulent shape, and reconnection of this magnetic turbulence provides the required conversion of magnetic into kinetic energy. Numerical computations of the field evolution  by \citet{Porth:2013tg} indeed demonstrate that the shocked toroidal field is (kink-)unstable, and acquires a turbulent aspect (see Figure \ref{fig:19}). It should, however, be mentioned that these authors use a large numerical resistivity (cell-size $2 \cdot 10^{16}$ cm!) to dissipate  kinks of mixed polarities, and thereby reduce  $\sigma$.

\subsection{Conclusions for radio pulsar winds}
\label{conclusion1}

The key problem in pulsars - and in cosmic magnetic structures in general - is to explain how magnetic energy is transported away from a dynamo to a conversion/dissipation region where energetic particles, radiation and outflows are observed. We have described  our present understanding of pulsar winds by focussing on their electric current systems. 

We have pointed out, on the basis of simple arguments, that an oblique magnetic rotator emits both an MHD wind and an embedded TEM, both of which are initially dominated by the magnetic field so that $\sigma >>1$. Both the wind and the TEM wave carry off angular momentum and energy away from the star in the form of Poynting flux. The electric current system has three domains: at the base it runs through the surface layers of the star across its magnetic field, thereby exerting a decelerating torque on the rotating star; next there is a largely force-free domain of the superfast wind where the currents follow the magnetic field lines; finally, the Poynting flux is converted into kinetic energy, so that $\sigma <<1$.  This can happen by a combination of effects:
\begin{itemize}
\item partly by inertial focussing towards the rotation axis and subsequent bulk acceleration in polar jets. However, since the pulsar is a slow rotator this does not happen unless important differences exist between the geometries of the polar winds of an oblique versus an aligned rotator;
\item partly by current starvation, the subsequent evolution of parallel electric fields, direct particle acceleration, current closure and bulk acceleration by the Lorentz force in a process which has been termed Generalized Magnetic Reconnection;
\item and partly by (CPFA) particle acceleration in the strong TEM.
\end{itemize}
To what degree `common' magnetic reconnection \citep{Parker:1957} and tearing play a role is not clear. We expect it to be confined to equatorial regions, and it may well be that its role has been severely overestimated. We like to point out that similar singular current layers, which exist in the solar corona between regions of different polarity, are not per se characteristic sites of solar flares. 

We wish to emphasize the twofold effects of parallel electric fields: First, they develop naturally in a current circuit with local or temporal shortages of charge carriers, where they lead to direct particle acceleration. An important example is the polar gap in a pulsar, which leads to abundant pair creation. Secondly, they also act as regions of `rupture'  on both sides of which ideal, force-free magnetic structures slide along each other (`slippage'), and where electric currents cross the magnetic field, thus leading to bulk acceleration by the Lorentz force. The example here is the region of current starvation far out in the pulsar wind.

Further, we have argued that singular current layers are often assumed to be present for convenience whereas consideration of realistic distributed (`body'-)currents leads to a better understanding of the physics of the current circuit and its dissipation. 

In the numerical computations it remains unclear whether  the computed macrostructures are insensitive to details of resistivity and viscosity, in particular the effects of numerical resistivity. This is a difficult question to answer because of the large range of length scales involved. For the radio pulsar the length scale varies over 11 decades, from the stellar radius of order 10 km, via the distance of the light cylinder $r_{LC} \equiv c/\Omega_\star \sim 1.58 \cdot 10^3 \;\;P_{0.033}$ km  ($P_{0.033}$ is the pulsar period in units of $0.033$ sec, the period of the Crab pulsar) to the distance of the reverse shock 0.1 pc. For jets in Active Galaxies it varies over 9 decades, from a few Schwarzschild radii $R_S= (2GM_{BH}/c^2)^{0.5} \sim 3 \cdot 10^{14} \;\;M_{BH}/M_\odot$~cm to the distances of knots and shocks at $10^{25}$ cm ($M_{BH}$ is the black hole mass).

\section{Current circuits in solar flares}\label{sect:solar}
In a solar flare, energy previously stored in a coronal current system is suddenly released and converted into heat, the kinetic energy of large numbers of non-thermal particles, and the mechanical energy of magnetized plasma set into motion. Broadly speaking, the heat and the excess radiation resulting from the non-thermal particles and their interactions with their environment constitutes the {\it flare}, whereas the motion of the magnetized plasma constitutes a {\it coronal mass ejection} (CME) which is often, but not always associated with a flare. The overall duration of the primary energy release phase (the impulsive phase) is on the order of 10 minutes, with significant variations in the energy release rate indicated by X-ray `elementary flare bursts' on $\sim$ 10 s timescales (and also lower-amplitude variations on shorter scales). The broadly accepted view of a solar flare is that the magnetic field is permitted to reconfigure by reconnection of the magnetic field - a highly localized process which nonetheless allows the global field/current system to relax to a very different energy state. The magnetic field changes dramatically in the impulsive phase, and steady-state models are certainly inappropriate.  It was for a long time assumed that the photospheric magnetic field did not change during a solar flare meaning that the photospheric current also did not change, but this is not correct. Substantial non-reversing changes in the photospheric line-of-sight and vector fields are observed to coincide with, and occur over, the few minutes of the impulsive phase of strong flares \citep{2005ApJ...635..647S,2012ApJ...759...50P}.

The usual approach in solar physics is to consider the corona to be an ideal MHD plasma (see (\ref{eq:2a})) in which the primary variables are plasma velocity and magnetic induction (magnetic field), $\vec v$ and $\vec B$, with current density $\vec j$ and electric field $\vec E$ being derived quantities. Of course, a flare can only happen when the ideal approximation breaks down, allowing field dissipation, and reconfiguration. The approach based on $\vec v$ and $\vec B$ as primary variables is in part because what is most readily \emph{observed} or at least inferred (e.g. from looking at EUV images of the solar corona to give an idea of projected field direction). One or more components of the magnetic field can be deduced from spectropolarimetric observations, while plasma flows can also be deduced spectroscopically or from feature tracking. Explaining field and flow observations has driven very successful developments in large-scale solar MHD models. However, one result of this is that the current circuit and how it changes is not much discussed at present.  Simple current circuit models for flares have long existed \citep[e.g.][]{1967SoPh....1..220A,1978ApJ...221.1068C}, but we do not often grapple with the reality of the time-dependent driving, re-routing and dissipation of currents in an intrinsically complex magnetic structure. 

This section overviews the several ways that electrical currents are considered in the context of solar flares, primarily as a means for storing energy in the magnetic field, and several models for the onset of a flare involve the classical instabilities of a current-carrying loop. The release of energy in a single current-carrying loop or a pair of interacting loops illustrates the transformation of non-potential magnetic energy to kinetic energy of accelerated particles, and to a mechanical and Poynting flux as the twist redistributes. However currents are also associated with \emph{singular magnetic structures} in a more complex coronal field geometry, such as \emph{X-lines}, \emph{current sheets} and \emph{separatrix surfaces}, and when reconnection occurs here fast particles can be accelerated forming a non-thermal particle beam, also constituting a current. More generally, the transport of the energy of a flare is often attributed to a beam of electrons which flows together with a counter-streaming population in a beam-return current system. 

\subsection{Coronal currents and energy storage}\label{sect:solar-energy}
\begin{figure}
\begin{centering}
\includegraphics[width=0.45\textwidth]{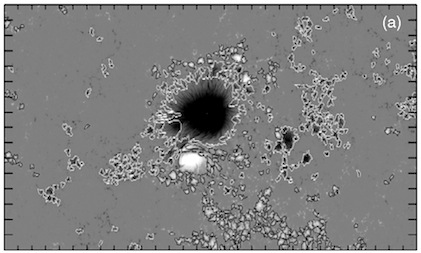}
\includegraphics[width=0.45\textwidth]{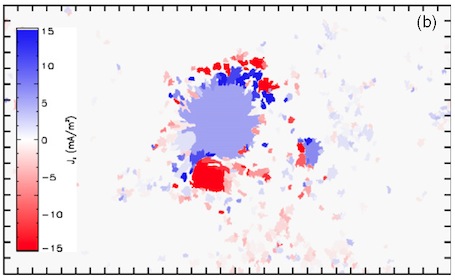}
\caption{Vertical magnetic field in magnetic active region (NOAA AR 10930) partitioned into flux elements (LH panel; black/white for opposite polarities) and showing calculated current density (RH panel). Non-neutralized currents tend to flow in strong field close to magnetic polarity inversion lines. The axes are solar $x$ and $y$, and tickmarks are 10'' apart - see \protect\citet{2012ApJ...761...61G} for detail; reproduced by permission of the AAS.}
\label{fig:georgoulis_currents}    
\end{centering}  
\end{figure}

The energy for a solar flare - i.e. the current system - is thought to arrive in the corona primarily by emerging through the photosphere already embedded in the twisted and stretched field of a magnetic active region \citep[e.g.][]{1996ApJ...462..547L,2007ApJ...671.1022J}. Small-scale photospheric stressing following emergence may also occur and be important in destabilization of a current-carrying coronal structure but it is not expected to be  energetically significant. Magnetic active regions are composed of a small number of dominant magnetic bipoles (quadrupolar fields are commonly observed in complex, flare-productive regions) and the strongest magnetic poles are associated with sunspots. The best data available at present demonstrate significant non-neutralized currents in strong field regions, meaning that at photospheric heights there are net currents entering the corona on one side of the polarity inversion line, and re-entering the photosphere on the other side. In particular, this is the case close to the active region magnetic polarity inversion line \citep[see e.g.][]{2000ApJ...532..616W, 2012ApJ...761...61G}. The sign of these currents is strongly correlated with the sign of their host magnetic element, suggesting a dominant sense of \emph{current helicity}. Overall the current is balanced within observational errors, in the sense that currents emerging on one side of the polarity inversion line have, somewhere, an oppositely-directed counterpart submerging on the other side. 
This non-neutralized current flow is an observed property of the active region on scales of tens of arcseconds, or thousands to tens of thousands of kilometers, i.e.  both the separation and the size of the main current and magnetic sources in the region. Figure~\ref{fig:georgoulis_currents} shows the magnetogram and net electrical currents calculated by \cite{2012ApJ...761...61G} flowing in individual magnetic fragments of the active region that hosted the last large (X-class) flare of Solar Cycle 23.  The size of the current sources, as illustrated in this figure may at least conceptually be linked to a scale of sub-photospheric generation or aggregation of ropes or bundles of magnetic flux carrying like (i.e. similarly directed) currents. The character of the sub-photospheric magnetic dynamo is of course not yet known, but  both turbulent convective flows and ordered shear and rotational flows are likely to play a role \citep[e.g.][]{2013Natur.497..463T} and thus a bundle with both a strong net current flow direction, and also a fragmentary character are to be expected. Oppositely-flowing currents already present in the sub-photospheric structures, and maintained during rise into the photosphere are hinted at by the interspersed red and blue sources in this figure, on the scale of a few arcseconds, though some of these  current sources are weak and are not considered as being significantly non-neutral by the authors.  The net currents in the largest negative and positive polarities in Figure~\ref{fig:georgoulis_currents} have values of $45 \times 10^{11}$ A and $-35 \times 10^{11}$ A respectively. The imbalance indicates that the main negative polarity must also provide current to sources other than the main positive polarity, as would be revealed by field reconstructions. Sub-resolution structures are also to be expected, and these may - indeed should -  include coronal return currents in thin surface sheets, providing partial neutralization of currents flowing through the photospheric boundary. 

A magnetic active region is generally more complicated than just a single twisted `rope' of magnetic flux carrying current from one sunspot to another, and magnetic field extrapolations based on the observed photospheric vector magnetograms are used to determine the 3D structure of the coronal magnetic field $\vec B$. Coronal electrical currents are obtained from the curl of this field. 
\begin{equation}
\vec j = \frac{c}{4 \pi} \nabla \times \vec B. 
\end{equation}
An underlying assumption is that the magnetic field is force-free, which means that electrical currents are aligned with the magnetic field, i.e. $\vec j \times \vec B/c = 0$. By definition, this assumption must break down locally in a flare when the current has to re-route and therefore at some location must cross from its pre-flare magnetic field to its post-flare magnetic field. The parameter quantifying the current is the $\alpha$-parameter which is constant along a given field line but may vary from magnetic field line to field line, i.e. from point to point across the magnetic source surface: 
\begin{equation}
\vec j = \alpha \vec B.   
\end{equation}
The case of $\alpha = 0$ corresponds to no current flowing and no free energy in the system: this minimum energy state is called the potential state. If $\alpha = const$ the field is called a linear force-free field, and if $\alpha$ varies from field line to field line the field is a non-linear force-free field (NLFFF). The  $\alpha$ parameter is calculated at the photosphere, and each photospheric location paired up with another location having the same $\alpha$, defining the \emph{magnetic connectivity} in a region. Magnetic field extrapolations are still somewhat of a `dark art', beset with difficulties (in particular in resolving the 180$^\circ$ ambiguity in the measured in-the-plane component of $\vec B$) but good agreement is found between analytic non-linear force-free fields and their reconstructed counterparts particularly in strong field, strong current locations \citep{2006SoPh..235..161S}. 
The mathematical formulation of magnetic field extrapolations, with field lines having at least one end in a photospheric source, means that it is not possible to find field-aligned currents closing completely within the corona. However, one could envisage energy-storing force-free `tokamak' structures entirely disconnected from the photosphere yet prevented from leaving the corona by overlying field, or (perhaps rather implausibly) by being threaded by an anchored magnetic loop. 
We note also that non-force-free structures in which $\vec j \times \vec B/c$ is balanced by a gas pressure gradient are thought not likely to be significant in flares, because coronal gas pressure gradients are weak, except perhaps in the neighbourhood of a dense coronal filament.

\begin{figure}
\begin{centering}
\includegraphics[width=0.6\textwidth]{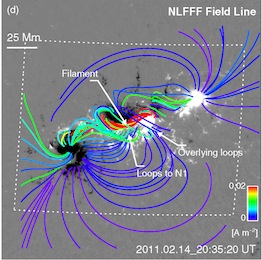}
\caption{The results of a non-linear force-free field extrapolation from the magnetic field observed around 5 hours before a major flare on February 15 2012. The lines are extrapolated field lines color coded by the calculated current; highest currents are concentrated in highly sheared and twisted magnetic field lying close to parallel to the polarity inversion line. From \protect\citet{2012ApJ...748...77S}, reproduced by permission of the AAS.}
\label{fig:sunetal_currents}    
\end{centering}  
\end{figure}

An example of the coronal currents deduced before a major flare on 15th February 2011 is shown in Figure~\ref{fig:sunetal_currents}. It indicates the variety of  different interconnected current systems that can arise and that can be resolved with NLFFF extrapolation techniques. If extrapolations are carried out before and after a flare, generally speaking the calculated currents in the flare region have changed. Depending on the time resolution available, observations do not always show that the magnetic energy represented by the currents has decreased (as must nonetheless happen.) However the flare may be a relatively small and transient perturbation on a system in which the overall magnetic energy is increasing.

Magnetic extrapolation, and MHD modeling of evolving active regions, calculates the force-free coronal current but does not address the current closure;  in these models, current emerges from one source on the photosphere and disappears into a sink.  This is adequate for most purposes, but in reality currents must close. This is presumed to happen somewhere below the photospheric layer (at least in conditions of slow evolution) in a layer where flux tube identity is established, and yet the flux tubes themselves are small enough that they can be stressed by plasma flow \citep{1989GMS....54..219M}. This implies a plasma dense enough that it can provide the mechanical stresses on the field-carrying plasma to twist it up. \cite{1995ApJ...451..391M} argues that this may happen deep in the solar convection zone. However, during a solar flare the energy release and therefore the coronal current evolution is much faster than the timescales on which signals about changing magnetic fields can propagate (Alfv\'enically) out of or into the region below the photosphere. The evolving current system in a flare requires current re-routing across the field at higher layers, in processes possibly analogous to magnetospheric substorms.

\subsection{Instability of current-carrying flux ropes and flare onset}\label{sect:solar-erupt}
Prior to a flare and CME a coronal current system exists in a stable or metastable equilibrium. There are many theoretical models for the processes leading to the loss of equilibrium \cite[see e.g. a recent review by][]{2013arXiv1309.7329A}; we focus our discussion here around the role of the current. Roughly speaking instability is to be expected when the axial current becomes strong, such that the toroidal component of the field exceeds the poloidal component. This can be seen as consistent with the result by \cite{1984ApJ...283..349A} that the maximum energy above the potential field that can be stored in a 3D stressed force-free field with the same normal field at the boundary is of the same order as that of the potential field itself. 

\begin{figure}
\begin{centering}
\includegraphics[width=0.45\textwidth]{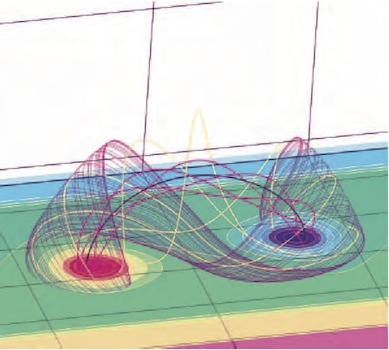}
\includegraphics[width=0.45\textwidth]{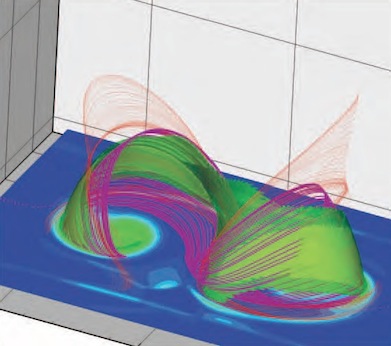}
\caption{Left panel: Sample field lines from an MHD simulation of a magnetic flux rope emerging into an overlying coronal arcade. Magnetic field contours are shown on the photospheric surface, with opposite polarities in magenta and blue colors. Field lines are shown for the flux rope and for the bald-patch separatrix surface. Right panel: associated current density shown by green photospheric color contours and coronal isocontours. From \protect\citet{2004ApJ...617..600G}, reproduced by permission of the AAS.}
\label{fig:gibsonetal_currents}    
\end{centering}  
\end{figure}

In one of the earliest models for an electrical current in the corona by \cite{1974A&A....31..189K}, an over-dense (straight) current-carrying filament in the corona is kept 
in stable equilibrium by a balance between three forces: the downward Lorentz force from the current acting on the background field of the overlying magnetic arch and the (relatively small) force of gravity are balanced by the upward Lorentz forces due to screening currents  induced above the photosphere by the filament's electrical current (also described mathematically in terms of a sub-photospheric `mirror current'). 
It was demonstrated by \cite{1978SoPh...59..115V} that an initially low-altitude current filament would become unstable and that the filament field structure and its entrained plasma would accelerate upwards if its current exceeded some critical value, or if it were given a sufficiently large upward displacement. If the filament is allowed to bend around to become a twisted loop, or `flux rope', anchored in the photosphere a semi-toroidal configuration arises and the destabilizing role of the curvature (or hoop) force of the loop on itself must be considered. An overlying dipolar field can keep this system in equilibrium. The resulting magnetic field of even a simple construction of this kind shows features - separatrix surfaces and concave-up field lines tangential to the photosphere (known as bald patches) - indicating the structural complexity present in realistic force-free fields \citep[e.g.][]{1999A&A...351..707T}. An example of the field and currents arising in an MHD model of this configuration is shown in Figure~\ref{fig:gibsonetal_currents}, from \cite{2004ApJ...617..600G}. On the left panel the current-carrying flux-rope emerges from one magnetic polarity (indicated by color contours on the surface) and enters another. color contours running parallel to the rope axis shows the feet of the overlying potential magnetic arcade. A selection of field lines are plotted for the twisted flux rope, and the 
concave-up bald-patch field. The bald-patch field lines are also plotted in the RH panel, and overlaid with current isocontours showing the current sheets set up at the interfaces between flux rope and arcade field. 

Numerical simulations can reveal details of the instability of current-carrying configurations, and how magnetic reconnection in the evolving field allows the instability to develop into a CME and flare. These simulations can exhibit quite baffling complexity; the results must be analyzed carefully to unravel the dynamic features that appear and disappear as instability and reconnection occurs. Reconnection takes place at the current sheets around, above and also below the flux rope, and this allows the kind of dramatic field reconfiguration necessary to extract energy from the field and launch a CME.

\subsection{Fixed circuit (single loop) flare models}\label{sect:solar-fixed}
Flare models can be much simpler than the elaborate MHD model shown above, their purpose being to explain the energy release, transport and conversion rather than dynamics. A basic flare structure is a simple, single loop (as is often seen in soft X-ray or extreme UV images) and models for flares invoking the enhanced dissipation of field-aligned electrical currents within such a loop have a long heritage, starting with \cite{1967SoPh....1..220A}. In this type of model an electrical current $I$ encounters a region of enhanced resistivity $R$. By analogy with various laboratory examples, Alfv\'en \& Carlqvist argue that - if the current cannot divert around it - the entire magnetic energy stored in the current system will tend to dissipate in the region of enhanced resistivity. The enhanced resistivity in the Alfv\'en \& Carlqvist model starts with a small local plasma density depletion. Persistence of the current across this region  means that the (more mobile) electrons must be accelerated into the depletion and decelerated as they exit it. The local parallel electrostatic electric field thus generated drives ions in the opposite direction, further enhancing the depletion of charge-carriers and creating an electrostatic double-layer. This requires that the current-carrying electrons move through the double layer faster than the electron thermal speed (otherwise the potential can be shorted by redistributing thermal electrons). It was noted by  \cite{1972ApJ...176..487S} that the situation where electrons drift faster than the local thermal speed is equivalent to the condition for the current-driven Buneman instability. The suggestion follows that the microinstability resulting from the Buneman instability provides the enhanced resistivity in the Alfv\'en \& Carlqvist model. The ion-acoustic instability has an even lower threshold for onset, occurring when the electrons and ions have a relative drift that is greater than the ion sound speed but requires the electrons to be much hotter than the ions. Both of these instabilities provide an anomalous resistivity by setting up waves which scatter electrons and ions. 
                                                                                                                                                                                                                                                                                                                                                                                                                                                                                                                                                                                                                                                                                                                                                                                                                                                                                                                                                                                                                                                                                                                                                                                                                                                                                                                                                                                                                                                                                                                                                                                                                                                                                                                                                                                                                                                                                                                                                                                                                                                                                                                                                                                                                                                                                                                                                                                                                                                                                                                                                                                                                                                                                                                                                                                                                                                                                                                                                                                                                                                                                                                                                                                                                                                                                                                                                                                                                                                                                                                                                                                                                                                                                                                                                                                                                                                                                                                                                                                                                                                                                                                                                                                                                                                                                                        
The power  $I^2 R$ dissipated in the resistive region produces flare heating, and the potential drop across the resistance may be able to accelerate charged particles. The typical power needed for a flare is $\sim 10^{27} - 10^{28}$ ergs~s$^{-1}$ ($10^{20} - 10^{21}$~W), and with a current of $\sim 10^{20}-10^{21}$ statamps ($3.3 \cdot 10^{10} - 3.3 \cdot 10^{11}$~A) this means that the resistance is around 
$10^{-13} -10^{-12}$~s~cm$^{-1}$  ($9 \cdot 10^{-1} - 9 \cdot 10^{-2}$~$\Omega$). This gives rise to a potential drop of around $3 \cdot 10^{10} - 3 \cdot 10^{9}$~V, producing electrons and protons up to about $10^{10}$ eV. This is very much larger than the kinetic energy of a few tens of keV typical of flare electrons, so a single potential drop in this system is not a good model for flare particle acceleration.

\begin{figure}
\begin{centering}
\includegraphics[width=0.6\textwidth]{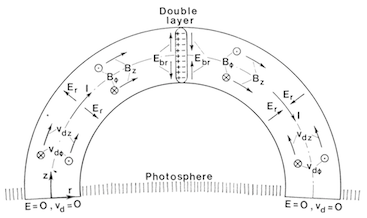}
\caption{The radial electrostatic field $E_{br}$ formed at the boundary of a double layer crossed into the magnetic field $\vec B = (0, B_\phi, B_z)$ gives rise to a plasma drift $\vec v = c \vec E \times \vec B / E^2=(0, v_{d\phi}, v_{dx})$, and a Poynting flux $S = c(\vec E \times \vec B)/4\pi$.  From \protect\cite{1979SoPh...63..353C}, \copyright 1979, D. Reidel Publishing Co.}
\label{fig:carlqvist_loop}
\end{centering}
\end{figure}

In a later iteration of this model, \cite{1979SoPh...63..353C} demonstrated how part of the energy, previously stored in the twisted field structure, arrives at the double layer as a Poynting flux. Figure~\ref{fig:carlqvist_loop} illustrates an electrostatic double layer introduced into a (twisted) coronal loop, leading to a radial electrostatic field on its boundary (due to its finite radial extent). When crossed into magnetic field of the current-carrying loop this electrostatic field gives rise to a  Poynting vector  $\vec S = c(\vec E \times \vec B)/4\pi$, and a plasma drift with the same direction ${\vec v}_d = c(\vec E \times \vec B)/B^2$, and therefore with both an axial and a toroidal component. The azimuthal component of the drift represents the unwinding of the twisted flux tube across the resistive region at the top of the loop. This idea, forming an interesting bridge between Alfv\'enic energy transport and particle acceleration, was subsequently generalized by \cite{1992ApJ...387..403M} to include an arbitrary source of resistance, which could in principle be placed anywhere in the current-carrying loop.

A potential drop can of course accelerate particles, but it is not straightforward in this case. For a flare, we require a large number of non-thermal electrons with energy of a few tens of keV. The single potential drop implied by the current and flare parameters would instead produce extremely high energy particles, but a small number of them (preserving the current $n_e q v$ across the resistance). A more sophisticated view is necessary. 

\subsection{Current diversion flare models}\label{sect:solar-diversion}
There is ample observational evidence for the involvement of multiple magnetic structures in a flare. The presence of more than two bright chromospheric (or occasionally photospheric) foot points during a flare demonstrates that multiple flux systems are present.  Quadrupolar systems, with at least 4 groups of foot point, are rather common. The spreading of those foot points or ribbons in the chromosphere that is particularly evident in H$\alpha$ and UV emission indicates that from instant to instant the identity of the bundle of magnetic flux that is involved in the flare is changing - in a process very similar in appearance to the racing of auroral curtains across the sky (see Section \ref{fa_current} and Figure \ref{fig:arcs}). The appearance of previously invisible flare loops indicates that the topology has  changed. This happens during the impulsive and the gradual phases of the flare; a particularly (perhaps the only) compelling solar observation of this was recently presented by \cite{2013NatPh...9..489S}, but the ample circumstantial evidence for reconnection means that its existence in the flare corona is not generally in doubt.  A large fraction of flares thus involve the formation of new flux and new current systems as magnetic connections are broken and reformed. 

\begin{figure}
\begin{centering}
\includegraphics[width=0.6\textwidth]{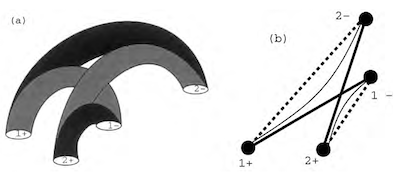}
\caption{Two current-carrying loops (grey) which reconnect to give the (black) post-flare configuration. Their coronal current paths are represented by thick solid and dashed lines respectively (with the thin solid line representing an intermediate stage during which the field-lines/current paths are shortening. From \protect\citet{1997ApJ...486..521M}, reproduced by permission of the AAS.}
\label{fig:melrose_loops}    
\end{centering}  
\end{figure}

\cite{1997ApJ...486..521M} associates the flare energy with the change in energy stored in coronal currents as the current path shortens, and this process was explored in a model of two interacting current-carrying loops, as sketched in Figure~\ref{fig:melrose_loops}. Reconnection was involved only in so far as it allowed the field to reconfigure and the currents to re-route and the field lines to shorten, corresponding to the release of energy. More recent work by \cite{2012ApJ...749...58M,2012ApJ...749...59M} demonstrates how this process happens; it is argued that the cross-field current enabling the re-routing of previously field-aligned currents in the corona is a result of  the intrinsically time-varying character of the flare magnetic field, which is quite appropriate for the flare impulsive phase. The resulting inductive electric field plays a critical role in the overall problem. The discussion below summarizes Melrose's arguments. 

The inductive electric field is given by 
\begin{equation}
\vec \nabla \times E_{ind} = - \frac{1}{c}\frac{\partial{\vec B}}{\partial t}. 
\end{equation}
A time-varying $\vec E_{ind}$ sets up a polarization drift - in this way the individual responses of the plasma particles to the changing magnetic field become important. This polarization drift is

\begin{equation}
\vec v_{drift} = \frac{mc^2}{q|\vec B|^2}\frac{d\vec E_{ind}}{dt}.
\label{eq:l30}
\end{equation}
The changing magnetic field thus leads to a  drift current (since the drift velocity depends on particle charge). It is usually assumed that the component of the inductive electric field along the field is neutralized by the flow of mobile electrons. The perpendicular component of this current is the polarization current 
carried mainly by the ions (see (\ref{eq:l30})) 
and flowing across the field, exactly as is required for current re-routing. It is related to the inductive electric field by

\begin{equation}
\vec j_{\perp} = \vec j_{pol} = \frac{c^2}{v_A^2}\frac{1}{4\pi}\frac{\partial \vec E_\perp}{\partial t}
\label{eq:l31}
\end{equation}
As can be seen from (\ref{eq:l30}) and (\ref{eq:l31}), the polarization current is proportional to the displacement current, but the latter is negligible in the solar case (unlike in the pulsar case, see Section \ref{obliquity},  (\ref{eq:j22}) and ({\ref{eq:j23})).  
Melrose also argues that in the flare as a whole the $\vec j_{pol} \times \vec B/c$ force that develops, once the change in magnetic field has been triggered by some external process, can drive field into the magnetic reconnection region, sustaining the process without any mechanical forces (i.e. pressure gradients) being necessary. 

Within an individual current-carrying (twisted) flux tube the $\vec j_{pol} \times \vec B/c$ force generates a rotational motion of the plasma, i.e. a propagating twist or torsional Alfv\'en wave that carries energy away (see previous Section). The idea that the energy in a flare can be transported by Alfv\'enic perturbations has also been explored by, e.g., \cite{2006SSRv..124..317H} and \cite{2008ApJ...675.1645F}. The reconnection region, in which the rapid field variations at the root of the process begin, can be arbitrarily small (with arbitrarily small energy dissipation locally); its role instead is to allow the transfer of current from pre-reconnection field to post-reconnection field, and in doing so it  launches Alfv\'enic Poynting flux along just-reconnected field lines, redistributing energy throughout the corona with the current. 

The discussion above indicates how the currents in the corona may be rerouted and in the process energy transported away from the reconnection site and a new current profile established throughout the corona. In the magnetosphere/ionosphere system the current restructuring implied by substorms (flare analogues) involves current closure in the partially ionized layers of the ionosphere, supported by the Pedersen resistivity (Section \ref{fa_current}). The deep chromosphere is also a partially ionized plasma, however the high density and resulting strong collisional coupling between ions and neutrals means that the collisional friction that provides the Pedersen resistivity is too small to support cross-field currents \citep[e.g.][]{2006SSRv..124..317H}.  That is not to say that ion-neutral collisions have no role to play in the flare energy dissipation; if the period of the Alfv\'enic pulse (moving the ions) is short compared to the ion-neutral collisional timescale then frictional damping can take place.  The ion-neutral collisional timescale is longest around the temperature-minimum region of the solar chromosphere where the number density of protons is also at a minimum, and significant wave damping can occur here for wave pulses of duration a second or less \citep{2013ApJ...765...81R}.

An Alfv\'en wave entering the chromosphere passes through a region in which the Alfv\'en speed is varying rapidly; the reflection and transmission and damping of the wave needs to be carefully calculated \citep[e.g.][]{2001ApJ...558..859D}. Moreover, if the wave enters at an angle to the density gradient, mode conversion can also occur \citep[e.g.][]{2006ApJ...653..739K}. Therefore, the wave energy is unlikely to be dissipated straightforwardly, in a ``single pass'' through the chromosphere or photosphere, but multiple reflections leading to counterstreaming waves, generation of plasma turbulence, and other complexities, are to be expected.
However, at some level the photospheric currents are fixed, and somehow the new and old currents must match, which implies current dissipation or current diversion into a singular layer.  
At some point in the atmosphere the field becomes line-tied, and the high inertia of the material means that the Alfv\'en wave front reaching the photosphere cannot continue.
Its energy and current must be dissipated in full or in part. Dissipation at this layer, as well as the damping occurring elsewhere in the atmosphere, must be what provides the intense heating and particle acceleration that characterize a solar flare.  In broad terms, the stressed magnetic field in the wave pulse leads to field-aligned electric fields which can be a source of both heating and acceleration in the plasmas of the lower atmosphere. 
This is described further in Section~\ref{sect:solar-acceleration}. Another effect of the high inertia of the photospheric material is that any excess photospheric current can be diverted across the magnetic field into a thin envelope around the coronal flaring region. It can do so since the photospheric material can take up the stresses resulting from the Lorentz force.

\subsection{Singular current-carrying structures in a complex corona}\label{sect:solar_qsl}
As indicated in Section~\ref{sect:solar-erupt} the magnetic field and the current structures arising from something even as simple as a current-carrying flux rope embedded in an overlying potential field arcade can be quite complex. The extrapolated magnetic field shown in Figure~\ref{fig:sunetal_currents} suggests multiple magnetic domains (regions with the same magnetic connectivity).

\begin{figure}
\begin{centering}
\includegraphics[width=0.6\textwidth]{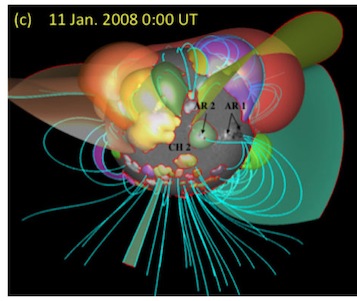}
\caption{A rendering of the separate coronal magnetic domains over the entire Sun (semi-transparent colored surfaces), and a few selected field lines (in cyan). The semi-transparent surfaces are in fact separatrix surfaces, separating regions of different magnetic connectivity. Part of the heliospheric current sheet (in semi-transparent yellow) is also shown. From \protect\citet{2012SoPh..281..237V}, \copyright 2012, Springer Science+Business Media B.V. .}
\label{fig:domains}    
\end{centering}  
\end{figure}

%
Coronal field configurations can be analyzed to identify sheet-like magnetic discontinuities - regions of rapid change of field-line connectivity - known as separatrices. These are regions where, in non-potential configurations, current sheets can develop \citep[e.g.][]{1988ApJ...324..574L}. In fields determined from realistic magnetic boundaries singular separatrices are not found; the relevant structures are instead quasi-separatrix layers (QSLs). An example of this for the whole Sun field is shown in Figure~\ref{fig:domains}, where the semi-transparent separatrix domes indicate the interfaces between fields which are closed on the scale of an active region, from those which open to the heliosphere.

It is now well-known that, in flaring regions, the intersection of QSLs with the photosphere corresponds closely with sites of strong flare radiation, i.e. the flare ribbons  \citep[e.g.][]{1997A&A...325..305D} (this is discussed in more detail in Cargill et al. (this volume). 
The flare ribbons - or at least the subset of them defined by the non-thermal X-ray and optical foot points - are where the majority of the flare energy arrives at the lower atmosphere. However, the field extrapolations suggest that the bulk currents are stored in and released from large-scale flux ropes. It is not clear how these pictures match up - whether in fact sufficient current can be stored in QSLs as opposed to flux ropes, or whether the process of flux-rope destabilization involves reconnection with and current transfer onto the QSL field. 

A further type of current system often discussed in the solar flare context is that which exists in the dissipation region of a reconnecting structure, where the magnetic field is sufficiently small and thus the particle Larmor radii are large compared to the scale length on which the magnetic field varies. Reconnection - even in the steady state - involves the advection of magnetized plasma into a current sheet or X-line. This results in a $-\vec v \times \vec B/c$ electric field, and electrons which are demagnetized in a small region around the X-point can be accelerated in the direction of this electric field. Ions, due to their larger gyro radii, are demagnetized over a larger region and are accelerated in the opposite direction. 
Where this current closes will depend on the configuration of the magnetic field and on the orbits of the particles; these exhibit properties of dynamical chaos meaning that they can be deflected out along the magnetic separatrices into higher field regions where they become re-magnetized and may leave the system, or mirror and return for further acceleration \citep{2006SoPh..236...59H}. In 2.5D the X-line or current sheet extends indefinitely and currents run uniformly along it, but of course in 3D these currents must close; the X-line is generalized as a magnetic separator which meets the photosphere at either end - the model of \cite{1986A&A...163..210S} shown in Figure~\ref{fig:somov_rainbow} indicates one possible geometry for this.

\begin{figure}
\begin{centering}
\includegraphics[width=0.6\textwidth]{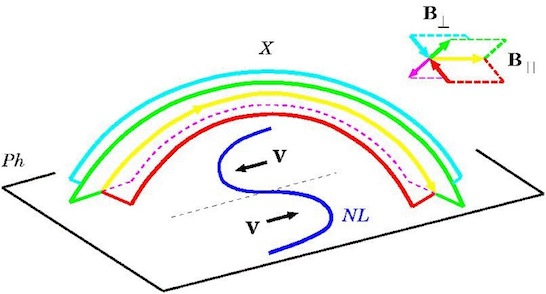}
\caption{The X-line, indicated by the field vectors in the top right, is the separator between reconnecting magnetic domains. In this cartoon it is projected from one photospheric endpoint to the other, offering a current path. In a localized region such as extracted in the top right, the reconnection electric field flows parallel to the X-line. Credit: \protect\citet{1986A&A...163..210S}, reproduced with permission  \copyright ESO.}
\label{fig:somov_rainbow}    
\end{centering}  
\end{figure}

\subsection{Beams and return currents}\label{sect:solar-return}
Non-thermal particles are a defining property of the majority of solar flares, and explaining where and how they are accelerated is a problem of long standing. The standard model of solar flares separates the region where particles are accelerated from the region where the flare energy is dissipated as radiation, placing the former in the corona close to (or perhaps in) the magnetic reconnection region and the latter in the lower atmosphere - the chromosphere. The chromosphere is where the flare radiation is primarily generated and where the signatures of accelerated particles are primarily seen. If they are accelerated in the corona the particles must  then travel from the corona to the chromosphere transporting energy, and must therefore have a significant component of velocity directed along the magnetic field, in a beam-like distribution. 

The properties of the accelerated particles can be deduced from the radiation that they produce - hard X-rays (at typically a few 10s of keV) for electrons, and $\gamma$-rays for ions. Most of the work on flare non-thermal particles focuses on electrons, simply because HXRs are more readily observed than are $\gamma$-rays, and our knowledge of ions is thus rather poor. In particular, we do not know whether electrons and ions are always accelerated together in flares, but electron acceleration is a central feature. Energy transported to the chromosphere by a beam of electrons must be dissipated there, which means that the electrons give up their energy and stop; at the minimum this involves Coulomb collisions in the dense chromospheric plasma. Under this assumption, the so-called \emph{`collisional thick-target model'},  the rate of incident electrons required to explain the observed HXRs can be calculated \citep{1971SoPh...18..489B}. The deduced rate of $\sim 10^{35} - 10^{36}$ electrons s$^{-1}$ would, if not neutralised, constitute a colossal current, of $10^{14}-10^{15}$ A ($3\cdot 10^{23}-3\cdot 10^{24}$ statamp), orders of magnitude greater than is present in the pre-flare corona (and with a correspondingly large associated magnetic field). However, this current will not exist un-neutralized. Driven by the inductive electric field of the beam and the (transient) electrostatic field at its head, a return current of electrons will flow \citep[e.g.][]{1990A&A...234..496V}, drawn presumably from the chromosphere where there is a large charge reservoir. It has frequently been proposed that this return current could replenish the accelerator, thus neatly solving the problem of how to replenish the acceleration region (which operates in a tenuous corona, and cannot pump out $10^{35}$ electrons per second for long). It remains to be seen what the overall electron current circuit joining a chromospheric foot point and the coronal acceleration region would look like. In the case of a free-streaming beam, a return-current could flow co-spatially with the beam. Such a return current would not be able to re-enter an accelerator that involves a large-scale direct electric field; in other DC field accelerators such as around a reconnecting X-line the return current could re-enter the accelerator if particles could drift across the field (similarly in accelerators with many small-scale and possible transient current sheets). However, there are doubts about whether the intense electron beam demanded by observations \citep[e.g.][]{1976SoPh...48..197H, 2011ApJ...739...96K} can co-exist stably with its co-spatial counter-streaming return current, or instead have its energy substantially dissipated as heat via an instability.  Other configurations with a large-scale DC electric field, and thus without a co-spatial return current, have also been discussed. For example in the sketch from \cite{1991ApJ...375..382W} shown in Figure~\ref{fig:winglee_returncurrents}) the return currents flow over the surface of the channel through which the electron beam is accelerated and flows, and ions are drawn into the channel across the magnetic field, and are themselves accelerated (in the opposite direction to the electrons).

\begin{figure}
\begin{centering}
\includegraphics[width=0.6\textwidth]{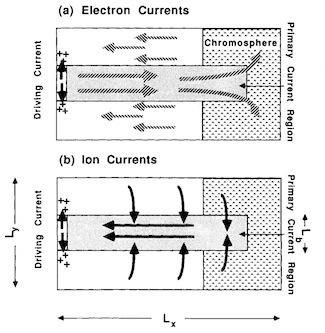}
\caption{Schematic of ion and electron currents produced by a cross-field driving current in the corona, and closing across the field in the lower atmosphere. From \protect\citet{1991ApJ...375..382W}, reproduced by permission of the AAS. In the upper panel, the electrons stream down the central channel into the chromosphere, and chromospheric electrons are drawn back up in external channels. In  the lower panel the ions, which are less strongly magnetized than the electrons, drift across the field to provide partial current closure in the beam channel, and once there are accelerated by the same electric field that accelerated the electrons.}
\label{fig:winglee_returncurrents}    
\end{centering}  
\end{figure}

An alternative is that a co-spatial beam of electrons and ions flows together through the corona in a neutral beam \citep{1990ApJS...73..333M} carrying zero net current. The ions would carry the bulk of the energy, and it is not entirely clear that such a beam arriving at the chromosphere can readily produce the non-thermal HXR spectrum observed though it is plausible under certain conditions \citep{2000A&A...353..729K}. It would be premature to discount this model, though there is  strong evidence for ions in only a few flares.

It should be emphasized that the main challenge in the electron-beam model is meeting the requirement of a high electron flux from the corona implied by HXR observations interpreted in the collisional thick-target model. Electrons decelerate collisionally in the chromosphere and only have a finite radiating lifetime (thus a finite and well-specified amount of radiation they can produce) before they join the thermal population. Electron flux requirements can be alleviated by allowing the electrons to be re-accelerated once they have entered the chromosphere \citep{2009A&A...508..993B}, e.g. in multiple current sheets or turbulence. However, the flare {\it energy} must still reach the chromosphere; the quiescent magnetic energy density of the chromosphere is not sufficient to explain a flare (e.g., compare the energy flux required to heat the chromosphere, of $\sim 10^3 - 10^4\;{\rm W~m^{-2}}$ ($10^6-10^7$ erg cm$^{-2}$ s$^{-1}$), with the flare requirement  of $\sim 10^{8}\;{\rm W~m^{-2}}$ ($10^{11}$ erg cm$^{-2}$ s$^{-1}$)). Flare energy may also be transported through the atmosphere by conduction (slow) or by MHD waves. The latter is an integral part of the holistic view of coronal currents discussed in Section~\ref{sect:solar-diversion}.

\subsection{Current circuits and flare particle acceleration}\label{sect:solar-acceleration}
A large fraction - estimates suggest up to 50\% - of the flare energy emerges in the kinetic energy of mildly relativistic electrons and ions. Electrons, which are more straightforward to detect via their bremsstrahlung radiation, typically have non-thermal energies of a few tens of keV. Spectra suggest that a true non-thermal tail can be distinguished from a core thermal distribution at energies exceeding $\sim$ 25~keV, and the spectral distribution is a power-law in energy $F(E) \sim E^{-\delta}$ with spectral index $\delta$ typically between 3 and 8. Imaging demonstrates that hard X-ray radiation, which is collisional bremsstrahlung from the non-thermal electrons, usually originates primarily in the dense chromosphere, except in rare `dense-loop flares' where the coronal density is high enough to lead to significant coronal non-thermal emission. In the flare standard model electrons are accelerated in the corona and propagate approximately collisionlessly to the chromosphere where they radiate. In this way the flare energy is also carried to the chromosphere, powering the optical and UV emission which is the flare's main radiation loss. As mentioned in Section \ref{sect:solar-return} the electron beam required may not be able to propagate stably together with its return current; this condition arises because the beam current density required by observations \citep[e.g.][]{2011ApJ...739...96K} is so high that it flows relative to its return current at speeds well in excess of the ion or even electron thermal speed, a condition that should lead to the ion-acoustic or Buneman instabilities respectively \citep{1977SoPh...52..117B}. Other ideas for flare particle acceleration, including placing the accelerator in or near the chromosphere, are being investigated \citep{2008ApJ...675.1645F,2009A&A...508..993B}, which may considerably reduce the number of accelerated electrons required (though not their total energy).

There are numerous models for particle acceleration in solar flares, and again we will be selective here, mentioning only those closely associated with the existence or disruption of electrical currents and thus providing a very incomplete view of flare particle acceleration. A recent overview of the many acceleration processes that may be operating in the solar atmosphere can be found in \cite{2011SSRv..159..357Z}. As mentioned in Section \ref{sect:solar_qsl} reconnecting structures such as current sheets or X-lines are a location where particle acceleration may take place, in the zone close to the singular region where one or more components of the magnetic field goes to zero and the particles become demagnetized and can accelerate freely in the electric field of the current sheet. Outside these regions particles are magnetized and the population as a whole experiences various kinds of drift, rather than the acceleration of a small tail to high energies. It is hard to see how a single macroscopic current sheet in the corona can fulfill the number requirement of $10^{36}\;{\rm electrons~s}^{-1}$ in a large flare; if the electrons are advected into the current sheet by the reconnection inflow at the external Alfv\'en speed 
%
$v_A = B/(4\pi n_e m_i)^{1/2}$ then the total electron flux that can pass through the sheet of area $A$ is $2 n_e v_A A \;{\rm electrons\;s}^{-1} = 
1.4 \times 10^{22} n_{9}B_{2} A$ for $n_{9}$ the electron number density in units of $10^{9}\rm{cm}^{-3}$ and $B_{2}$ the magnetic field strength in units of 100 G = $10^{-2}$T. For typical coronal values of $n_{9}$ = 1 and $B_{2} = 1$ the current area required is on the order of 10,000~km on a side (while being of the order of an ion skin-depth thick). Such a structure will not be stable and will fragment into many smaller structures. Fragmented current sheets distributed throughout a coronal volume may offer a better prospect. The need to continually supply the accelerating volume with fresh electrons remains the same though. For this reason, moving the current sheet accelerators into the low atmosphere - for example the transition region or upper chromosphere - is an attractive prospect since both the number density of electrons and the magnetic field strength are significantly increased; for example if $n_{9} = 100$ and $B_{2} = 10$, which might be found in the upper chromosphere, then $A$ is a more manageable 1000~km on a side, comparable in height to the thickness of the chromosphere. For acceleration, the electric field in the current sheet must exceed the Dreicer field, $E_D$ to allow significant electron runaway;

\begin{equation}
%
E_D = \frac{e \ln \Lambda}{\lambda_D^2} 
\end{equation}
where $\ln \Lambda$ is the Coulomb logarithm and $\lambda_D^2$ is the plasma Debye length. If we let the temperature in the upper chromosphere/transition region be $10^5 - 10^6$K then 
%
$E_D = 2.3 \cdot 10^{-5} - 2.3 \cdot 10^{-4} \;\rm{statvolt}\;\rm{cm^{-1}}\;=\;0.7 - 7\; \rm{V}\;\rm{m}^{-1}$.
If a parallel potential drop at least this large can be generated in a chromospheric/transition region current sheet, over a distance of a couple of thousand~km, this could potentially supply the required electron flux. 

Field-aligned acceleration by parallel electric fields is much more explored in pulsar and terrestrial magnetospheres than in solar flares; the literature dealing the obvious solar analogies is small. Field-aligned potential drops were part of the early current-interruption models of \cite{1967SoPh....1..220A} and \cite{1979SoPh...63..353C} mentioned in Section \ref{sect:solar-fixed}, but the basic problem with these is that for the inferred currents the parallel potential drops generated are of the order $10^{10}$ V, much higher than the typical electron energy. A more complex geometry is needed, and current fragmentation may provide the solution. \cite{1985ApJ...293..584H} suggested a very large number $n$ of filamented counter-streaming currents. If these currents provide the total power radiated by the flare, then they can do so at a lower potential drop per current channel. If the  potential drop along each current channel is reduced by a factor $n$ then the current per channel is increased to give the same power.
Potential drops of $\sim 100$~keV are obtained with $n \sim 10^5$ small currents, implying filamentation of the large-scale magnetic field into the same number of bundles \citep{1985ApJ...293..584H}. A model with many counter-streaming currents is in conflict with the observations of large net currents flowing in flare regions,
and the current channels would be very narrow. We know that most of the energy emerges in the hard X-ray part of the spectrum, and it is very hard to determine the size of HXR footpoints, but in well-observed flares scales on the order of a couple of arcseconds are found, and are consistent with optical source sizes on the order of $1-2 \times 10^{16}\rm{cm}^2$ \citep{2011ApJ...739...96K}. Each of the $10^5$  current channels would then have an area of $1-2 \times 10^{11}\rm{cm}^2$ at the footpoint. The coronal current channels could be larger, allowing for expansion of the field from the footpoint into the corona, but if the expansion of the flux tube is large then one must also incorporate the effect on particles of magnetic trapping.

In the magnetospheric case the existence of parallel electric fields is not in doubt. The exact altitude distribution of these fields is still debated \citep{Birn:2012}. The electric fields can be quite extended along the magnetic field in a quasi-static fashion, or they can be concentrated in multiple thin, so called double layers as for instance envisaged by \cite{1967SoPh....1..220A}. In the case of the magnetosphere, there seems to be no general agreement on the process generating the $E_{||}$ that is observed as also dispersive Alfv\'en waves have been observed and speculated as the source of small-scale structures in the aurora \citep{Birn:2012}. The solar case however, is much less well explored. 

Perhaps of most relevance here is that parallel electric fields in general play a role in dissipating magnetic stresses, as is required when the large-scale magnetic field relaxes. The general argument made by \cite{Haerendel:1994} and \cite{2012ApJ...749...59M} is that a parallel electric field allows the decoupling of two regions of magnetized plasma experiencing a shear flow relative to one another and perpendicular to the magnetic field threading both. In the solar case these would correspond to the photosphere where the magnetic field is frozen into a high-inertia fluid, and the corona through which an Alfv\'en wave pulse is traveling. Figure~\ref{fig:solar_haerendel94} sketches this process; in the left-hand panel, an initially vertical field (indicated by the solid lines with arrowheads), frozen in to the lower plasma, is distorted by a mechanical stress on the magnetized plasma at some (distant) top boundary.  In a zone termed the `fracture zone' by Haerendel, field reconfiguration takes place to release the stress, and the result is a plasma flow indicated in the Figure by $v_{\perp F}$. The `fracture' allows energy to be dissipated. Examining this scenario in cross-section, as in the right-hand panel of Figure~\ref{fig:solar_haerendel94}, the $v_{\perp F}$ in the region above the fracture results in an electric field $E_\perp$ perpendicular to the magnetic field (RH side of this panel). The $E_\perp$ is an electrostatic field, existing for much longer than the transit of the magnetic disturbance through the fracture zone, and is balanced by an oppositely-directed electrostatic field in the adjacent plasma (LH side of this panel), which sets that plasma into motion. The electrostatic equipotentials must be continuous, as sketched in the dashed lines, meaning that somewhere in the fracture zone there is an electric field parallel to the magnetic field, where acceleration may take place.  
The physical scales of the structure can be estimated based on the fact that in the region of parallel electric field in the fracture zone, the potential drop should be up to 10 MeV, to allow the acceleration of ions to $\gamma$-ray producing energies. The electric field should be in excess of the Dreicer field, which was earlier calculated as $0.7 - 7~{\rm V~m}^{-1}$ in the upper chromosphere, but - varying as it does with the plasma density - can easily reach $100~{\rm V~m}^{-1}$ in the mid-chromosphere. If we take $100~{\rm V~m}^{-1}$ as a plausible value, this means that to obtain MeV energies would require an accelerating length of around 100~km. The width of the equipotential structure shown in the right-hand panel of Figure~\ref{fig:solar_haerendel94} would then be on the order of 200~km. This is not dissimilar to the arcsecond-scale of flare ribbon widths.

The question of what provides the resistivity to support a parallel current  requires an examination of the microphysics, however in the presence of an external driver, the microphysics will have to find a way to keep up. So in principle the propagation of a magnetic shear (or twist) disturbance into a line-tied region can result in particle acceleration, in a region somewhere between the inertially line-tied photosphere and the frozen-in low-$\beta$ field of the corona. Initial evaluations by \cite{Haerendel:1994} concluded that for reasonable chromospheric parameters this process was capable of generating megavolt potential drops. Its application to the magnetosphere, the environment for which it was developed, is discussed in Section~\ref{fa_current}.

We remind the reader that a very similar conversion of energy has been proposed for the pulsar stellar wind in Section \ref{starvation}. Also there, parallel electric fields develop but now due to current starvation, particles are accelerated by these parallel electric fields, the outer wind slips over the inner fast rotating wind, the electric current closes, and leads to bulk acceleration (Generalized Magnetic Reconnection).  

\begin{figure}
\begin{centering}
\includegraphics[width=1.0\textwidth]{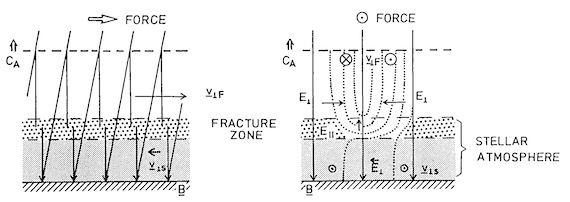}
\caption{View of a magnetic fracture zone, face-on (LH panel) and in cross-section (RH panel). See text for full description. From \protect\citet{Haerendel:1994}, reproduced by permission of the AAS.}
\label{fig:solar_haerendel94}    
\end{centering}  
\end{figure}

There are other possibilities for producing a parallel electric field, for example in a propagating Alfv\'en wave pulse traversing a low-$\beta$ plasma, such that the kinetic ($\beta \sim (m_e/m_p)^{1/2})$ or inertial ($\beta \sim (m_e/m_p)$)  regimes apply. In these cases a parallel field is sustained because the wave is traveling so fast that the finite thermal speed, and the finite inertia, of electrons cannot be ignored. Thus the electrons cannot instantaneously short out the $E_{||}$ that develops because of the cross-field polarization current arising from the induced electric field (see Section~\ref{sect:solar-diversion}).  The value of $E_{||}$ depends linearly on the parallel and perpendicular wavenumber of the wave, and in general needs small scales. This process has frequently been discussed in the magnetosphere/ionosphere context \cite[e.g. see review by][]{2000SSRv...92..423S}. It was examined by \cite{2008ApJ...675.1645F} for the solar analogue and found to be a viable process (i.e. generating an electric field in excess of the Dreicer field) in coronal densities and temperatures for waves with perpendicular scales of less than 5~km or so. Supplying adequate electrons from the tenuous corona to meet the HXR requirements remains problematic, but it is also possible that as the chromosphere heats during the flare onset and the Dreicer field decreases the effective acceleration in the $E_{||}$ of an inertial Alfv\'en wave becomes possible here too.

\subsection{Conclusions for solar flares}\label{sect:solar-concusion}
The complicated, current-bearing magnetic fields in solar active regions hosting flares must relax and release their energy in such a way as to inject it into the kinetic energy of non-thermal particles, which carry the bulk of a flare's energy which is then dissipated as heat. This may happen directly in the solar corona resulting in an intense beam of electrons - a current - propagating through the corona and balanced by a return flow of ambient electrons in a beam-return current system. Or, in processes more analogous with the magnetosphere, the relaxation might proceed in the form of Alfv\'enic perturbations launched by the introduction of an interruption to the current flow in a single loop, or by reconnection  in a system of two or more interacting loops. In the case of two interacting loops it may be shown how the inductive electric field leads to cross-field currents which allow the currents to reroute during reconnection, and drives a Poynting flux along the reconnected field. However we know that coronal magnetic topology is often more complicated than two interacting loops and this must be included in our thinking. Electrical currents with high individual particle energies can also be formed in the diffusion region of a reconnecting coronal structure, where individual particles become decoupled from the reconnecting components of the field and are accelerated in the 
$- \vec v \times \vec B/c$ field. However we note that \cite{2012ApJ...756..192L}, who treat the redistribution of energy stored in a current sheet in 2D, demonstrate that the vast majority of the stored energy is carried away from the sheet by fast magnetosonic waves launched by unbalanced Lorentz forces in the sheet, leaving little to be dissipated locally. They also comment that in 3D Alfv\'enic waves running along the separators will instead result. It remains to be seen how these different views of current-carrying structures can be reconciled, also with the complex current sheets that arise and that are observed to be strongly associated with the chromospheric locations of energy deposition.

An energy flow carried by magnetic perturbations must be dissipated somewhere, and there are several avenues through which it could lead to both heating and particle acceleration (though observations dictate that eventually non-thermal particle kinetic energy must dominate). Assuming that it is large enough to overcome Coulomb collisional friction, the parallel electric field set up by the (inertial) Alfv\'en wave propagating through the very low $\beta$ environment of the flare corona and upper transition region may accelerate or heat particles, depending on the ambient plasma conditions. The release of shear stresses in the wave as it travels towards the deep atmosphere where it is line-tied also generates a parallel field. More complex processes such as partial wave reflection in a converging field and the development of a turbulent cascade may also provide a source of first heating and then, as the plasma heats and becomes less collisional, particle acceleration in the lower atmosphere. The idea of waves rather than particles carrying flare energy through the corona to the chromosphere finds close analogy with the terrestrial magnetosphere-ionosphere system described in the next Section, and deserves further examination.

\section{Electric currents in the Earth's magnetosphere and ionosphere} \label{earthcurrent}

The main difference between electric currents around pulsars or in solar flares and currents in the Earth's magnetosphere and ionosphere is that the former are dominated by the effects coming from inside the pulsar or the Sun while the latter are generated by the interaction of the Earth's magnetosphere with the solar wind. At Earth the plasma of the solar wind travels with about 400-600~km/s, is comprised of electrons and mainly protons of $10^5 - 10^6$ K, carries a magnetic field of 1-10 nT ($10^{-5} - 10^{-4}$~ G) with variable orientation, and has an average density of 1-10 particles/cm$^3$. The solar wind plasma interacts with the originally dipolar magnetic field of the Earth, with a terrestrial surface magnetic field of 25 - 65 $\mu$T (0.25 - 0.65 G), (and all other magnetized bodies in the solar system like Jupiter, Saturn, and Uranus \citep{Russell:2003}) whereby the Earth's magnetosphere is formed into a tear-drop shape with a compressed side facing the sun and a long tail extending anti-sunward \citep{Roelof:1993}.  

\begin{figure} 
\includegraphics[width=30pc]{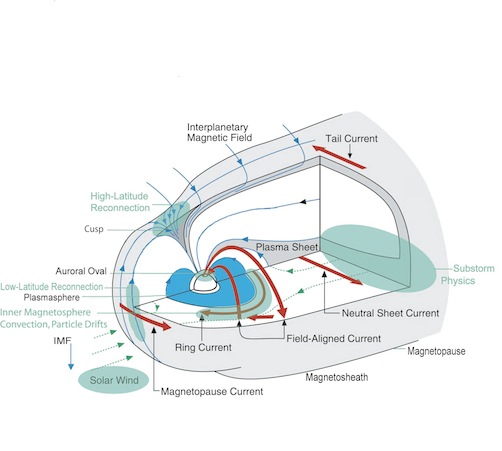}
\caption{Schematics of the Earth's magnetosphere with its major plasma regions, physical processes shown in green, and primary current systems shown in red. Modified from an original by C.T. Russell and R. Strangeway; from \protect\cite{Frey:2007}, \copyright American Geophysical Union.} \label{fig:magnetosphere}\end{figure}

\subsection{Magnetopause current} \label{mp_current}

Large-scale currents in space are the source of magnetic fields in distant regions and deform planetary dipole magnetic fields to produce planetary magnetospheres \citep{Parks:2004}. The prime example for this interaction is the generation of the magnetopause current, also called \emph{Chapman-Ferraro current}, and the subsequent deformation of the Earth's dipole field and generation of the magnetosphere (Figure \ref{fig:magnetosphere}). The magnetopause current is a typical example of a boundary current at the interface that separates the internal Earth's magnetic field and strongly magnetized, low density plasma (plasma $\beta < 0.1$ ) from the weakly magnetized, compressed and thus high-density ($\beta \approx 1$) plasma of the solar wind in the magnetosheath. It is generated by the solar wind protons and electrons that partially penetrate the geomagnetic field and spiral back out into the solar wind after half a gyration. The gyro radii of protons and electrons are very different and positive and negative particles gyrate in opposite directions which leads to the generation of the surface current. The magnetopause current responds to changes in the solar wind properties (especially pressure) and can quickly move the magnetopause which is very often seen in spacecraft data while crossing this space boundary (see e.g. \citet{Gosling:1967}).

The balance of external and internal forces at the magnetopause requires an equilibrium between the solar wind pressure $p_{sw}$, the magnetopause current $\vec j_{mp}$, and the Earth's magnetic field $\vec B$:
\begin{equation}
\nabla p_{sw} = \vec j_{mp} \times \vec B/c.
\end{equation}
This can be rewritten for the magnetopause current which for Earth runs from dawn to dusk \citep{Parks:2004}:
\begin{equation}
\vec j_{mp} = \vec B \times \frac{\nabla p_{sw}}{B^2}c.
\end{equation}
The magnetopause current roughly doubles the local magnetic field strength just inside the magnetopause and completely cancels it just outside. Besides particle signatures (change in density, temperature, and flow velocity) this is one indication in spacecraft data that the magnetopause has been crossed (e.g. \citet{Dunlop:2005}). A similar current should flow at the boundary between the solar system and interstellar space (heliopause), and at the interface region between a neutron star environment and intergalactic space, but so far has only directly been measured at Earth and Jupiter \citep{Kivelson:2000}.

In the equatorial plane the magnetopause current flows from dawn to dusk. In the magnetotail it splits into a northern and southern current, termed tail currents. The tail currents flow from dusk to dawn across the tail magnetopause and close to Earth they continue the dayside magnetopause current. Further down the tail the magnetopause currents close the neutral sheet currents that will be covered in the following section.

\subsection{Neutral sheet current} \label{ns_current}

The interaction between the solar wind and the Earth's magnetosphere gets substantially enhanced whenever the solar wind magnetic field has a southward component (negative $B_z$) and can connect itself to the Earth's dipole field in the subsolar region in the process named reconnection \citep{Parker:1957, Paschmann:2008}. Reconnected field lines are anchored on one end at the Earth's surface and the other end is carried by the solar wind flow downtail across the polar caps (Figure \ref{fig:convection}). One consequence of reconnection and plasma transport into the tail is the continuous supply of magnetic flux and fresh plasma towards the center of the magnetotail. Magnetic flux and plasma pressure would continuously increase if there were not a limiting process that converts magnetic field energy into heat and kinetic energy of particles and also removes plasma from the central magnetotail and that process is reconnection at the x-line in the tail.

\begin{figure}
\includegraphics[width=1\textwidth]{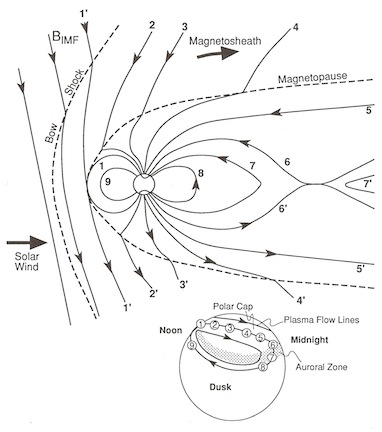}
\caption{Flow of plasma within the magnetosphere (convection) driven by magnetic reconnection at the dayside magnetopause. The numbered field lines show the succession of configurations a geomagnetic field line assumes after reconnection with an IMF field line (1), drag across the polar cap (2-5), reconnection at the x-line in the tail (6), ejection of plasma down the tail and into the solar wind (7'), and the subsequent return of the field line to the dayside at lower latitudes (7-9). From \protect\citet{Kivelson:1995}, \copyright Cambridge University Press, reprinted with permission.}
\label{fig:convection}      
\end{figure}

The magnetospheric tail at larger distances from Earth is separated into a northern and a southern lobe with opposite magnetic field directions, low density plasma ($<0.01$ cm$^{-3}$), and open magnetic field lines that extend well beyond the lunar orbit (see Figure \ref{fig:magnetosphere}). The flow of plasma towards the center of the magnetic tail creates the plasma sheet. This higher-density ($0.1-1$ cm$^{-3}$), hot ($T_i \sim$ 2-20 keV, $T_e\sim$ 0.4-4 keV) plasma region resides on closed magnetic field lines and contains plasma originating both from the solar wind and from the ionosphere \citep{Paschmann:2003}.  

The center of the plasma sheet is also called the neutral sheet, because the magnetic field direction reverses and the magnitude becomes very small ($<~5$~nT, $<5 \cdot 10^{-5}$~G). The screening of the magnetic field of different orientations is achieved by a diamagnetic current flowing across the magnetospheric tail from dawn to dusk \citep{Baumjohann:1996}. Its cause is the gradient in plasma pressure pointing from north to south in the upper half and from south to north in the lower (southern) half. An estimate of the neutral sheet current per unit length follows from
\begin{equation}
J = \frac{c B_T}{2 \pi}
\end{equation}
which separates the lobes and balances the tail magnetic field $B_T \approx 20$~nT which gives $J=30$~mA/m ($3\cdot 10^4$ statA/cm). Particles entering the neutral sheet from the northern or southern lobes become demagnetized and are transported toward dusk (positive ions) and towards dawn (negative electrons). Current continuity of the neutral sheet current is maintained through current closure with the tail magnetopause current in the north and south. If one could look down the tail from Earth and actually see the currents they would appear to form the Greek letter $\Theta$.

\subsection{Ring current}
\label{ring_current}

The ring current consists of ions trapped inside of $4-5 $ R$_E$ at the equator in the same region as the outer electron radiation belt, and their energy ranges from a few tens of keV up to a few MeV \citep{Paschmann:2003}. It is generated by the motion of trapped particles in an inhomogeneous planetary magnetic field as the particles undergo gradient and curvature drifts \citep{Parks:2004}. The gradient drift of a particle with charge $q$, mass $m$, perpendicular velocity $v_{\perp}$, and parallel velocity $v_{\parallel}$ is given by

\begin{equation}
\vec v_{\nabla B} = \frac{1}{2} \frac{m v^2_{\perp}}{q B} \frac{\vec B \times \nabla \vec B}{B^2} c,
\end{equation}
while the curvature drift of the particle is given by \citep{Parks:2004}

\begin{equation}
\vec v_{curv} = \frac{m v^2_{\parallel}}{q B} \frac{\vec B \times \nabla \vec B}{B^2} c.
\end{equation}
The action of both drifts will cause positive particles to drift westward in the Earth's dipole field, while negative particles drift eastward, creating a net westward current, the ring current. The effect of this westward current is a reduction of the local magnetic field at the surface of the Earth, especially close to the equator. The ring current becomes important during magnetic storms when the penetration of plasma from the tail drastically enhances the ring current and leads to a decrease of the magnetic field on the ground. During strong magnetic storms the depression of the equatorial magnetic field at the surface of the Earth can reach several hundred nT, or up to 1 \%. Defining $W$ as the total energy of the particles one can write the total current as 

\begin{equation}
%
I = - \frac{3W}{2 \pi r^2 B}c,
\end{equation} 
and the perturbation $\Delta \vec B_T$ of the Earth's magnetic field $B_s$ at the surface by this current can be expressed by 

\begin{equation}
%
\frac{\Delta \vec B_T}{B_s} = - \frac{2W}{B^2_s R^3_E} \vec z\end{equation}
The unit vector $\vec z$ points northward and the negative sign accounts for the reduction of the surface magnetic field by the westward flowing ring current.

\begin{figure}
\includegraphics[width=1\textwidth]{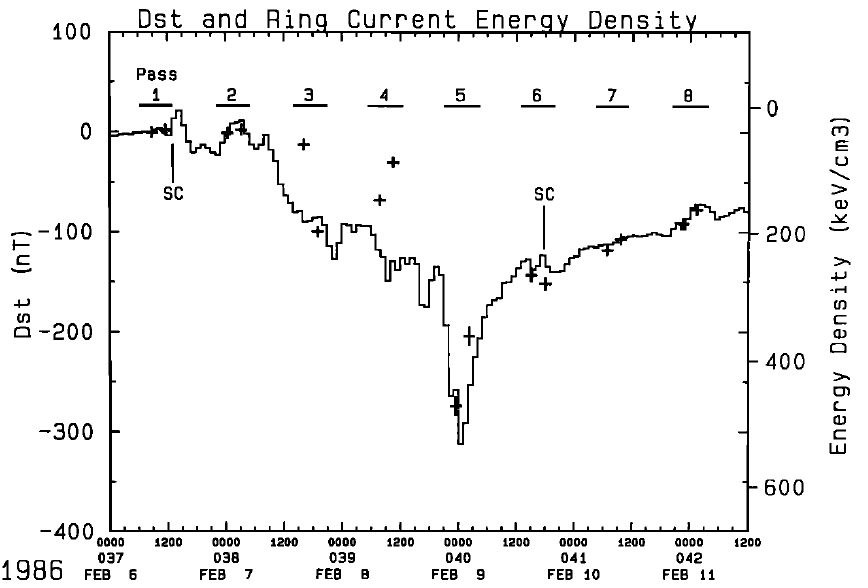}
\caption{Development of the Dst index during the large magnetic storm in February 1986 and the increase of the energy content of the ring current between L=3-5 (expressed in R$_E$). The line of Dst refers to the left scale, the plusses refer to the right scale, which is inverted to follow the Dst trends. Adaptation from \protect\citet{Hamilton:1988}, \copyright American Geophysical Union.}
\label{fig:ring_current}      
\end{figure}

Geomagnetic storms usually start with the impact of a strong and persistent southward interplanetary magnetic field together with an increased solar wind dynamic pressure on the dayside magnetopause. The enhanced pressure pushes the magnetopause inward by several Earth radii, enhances the magnetopause current, and temporarily causes an increase of the magnetic field on the ground. This phenomenon is known as the storm sudden commencement (SSC) and can be seen as the slight increase of the Dst (Disturbance storm time index, \citet{Mayaud:1980}) index in Figure \ref{fig:ring_current}. During the main phase of the magnetic storm charged particles in the near-Earth plasma sheet are energized and injected deeper into the inner magnetosphere drastically enhancing the quiet time ring current and producing the so-called storm time ring current. This leads to a strong decrease of the Dst index and marks the main phase of the storm (Figure \ref{fig:ring_current}). This main phase lasts generally for several hours but can last for several days during severe storms. After a turn of the IMF to northward directions the recovery phase of the storm begins during which the injection of plasma sheet material slows down or stops completely and various loss processes (primarily charge exchange and precipitation) weaken the ring current. The recovery phase can last several days up to several weeks.

During the early years of space research it was thought that the ring current consists only of protons. However, observations showed that a substantial fraction of the ions can be $O^+$ and $N^+$ \citep{Hamilton:1988}. During the example in Figure \ref{fig:ring_current} it was shown that more than 50 \% of the ring current ions were oxygen and nitrogen thus demonstrating the importance of the ionosphere as a potential source of ring current particles during major magnetic storms. Clues about how these ionospheric ions are accelerated and transported out into the magnetosphere come from correlated observations of electric field fluctuations and upward flowing ions. These observations indicate that low frequency  (frequency $<$ ion cyclotron frequency) electric field fluctuations with amplitudes of $\ge 0.1$ V/m accompany upward flowing ions. Several researchers have shown theoretically that large-amplitude electromagnetic waves can give rise to a significant ponderomotive force that can accelerate and transport ions out of the ionosphere and favor acceleration of heavier mass species. Thus, $O^+$ could be accelerated more effectively than $H^+$ ions, thereby explaining the dominance of oxygen species in the ring current during some major storms \citep{Parks:2004}.

Besides its importance on the ground magnetic disturbance during magnetic storms, the ring current also influences the size of the auroral oval \citep{Schulz:1997}. The storm time ring current stretches the magnetospheric magnetic field outward thus increasing the amount of open magnetic flux in the polar cap. As the inside of the auroral oval marks the boundary between open and closed magnetic field lines this increase in open flux enhances the overall size of the auroral oval by about $4^\circ$ for each -100 nT (-$10^{-3}$ G) of Dst.

\subsection{Field-aligned currents} \label{fa_current}

The electric fields that are associated with the convection process in the magnetosphere map along the magnetic field to the ionosphere where they create currents that are perpendicular to the magnetic field. In the terrestrial ionosphere both electrons and ions are scattered by collisions with neutrals and the generalized Ohm's law describing the relationship between the electric and magnetic fields and the current becomes: 

\begin{equation}
\vec j = \sigma_{\parallel} \vec E_{\parallel} + \sigma_P \vec E_{\perp} - \sigma_H (\vec E_{\perp} \times \vec B)/B
\end{equation} 
where the ionospheric Pedersen conductivity $\sigma_P$ governs the Pedersen current in the direction of that part of the electric field $\vec E_{\perp}$ which is transverse to the magnetic field. The Hall conductivity $\sigma_H$ determines the Hall current in the direction perpendicular to both the electric and magnetic field $-\vec E \times \vec B/c$. The parallel conductivity $\sigma_P$ governs the magnetic field-aligned current that is driven by the parallel electric field component $E_{\parallel}$ \citep{Baumjohann:1996}.

\begin{figure}
\includegraphics[width=1\textwidth]{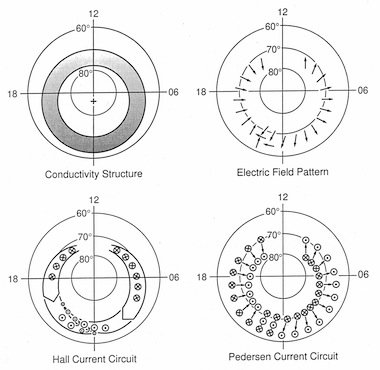}
\caption{Sketch of conductivity, electric field, and Hall and Pedersen currents in the auroral ionosphere. The dots show the upward and the crosses the downward field-aligned currents. From \protect\citet{Baumjohann:1996}, \copyright Imperial College Press.}
\label{fig:conductivity}      
\end{figure}

The large-scale field-aligned currents are traditionally divided into two classes: Region-1 and Region-2 currents. Region-1 currents flow at higher latitudes poleward of the Region-2 currents and close at the magnetospheric boundaries. Region-2 currents run at lower latitudes equatorward of the Region-1 currents and close in the inner magnetosphere. In the ionosphere these currents close through horizontal currents as shown in Figure \ref{fig:conductivity}. Magnetospheric convection determines the orientation of the currents with Region-1 flowing out of the ionosphere in the evening sector and into the ionosphere in the morning sector (see Figure \ref{fig:fac}). On the dayside these currents connect to the dayside magnetospheric boundaries (e.g. \citet{Janhunen:1997}). However, the nightside currents most likely originate from the plasma sheet \citep{Tsyganenko:1993}. The total current in this circuit is 1-2 MA ($3-6 \cdot 10^{15}$ statA) \citep{Paschmann:2003}.

The Region-2 currents flow equatorward of the Region-1 currents and in opposite directions. They flow into the ionosphere in the evening sector, and they flow out of the ionosphere in the morning sector (see Figure \ref{fig:fac}). At high altitudes these currents merge with the ring current in the inner magnetosphere. The total current in the Region-2 circuit is slightly smaller than in the Region-1 current circuit, typically less than 1 MA. The slight overlap of the two regions with the reversal of current directions shortly before midnight is called the Harang discontinuity \citep{Koskinen:1995} which may play an important role in the localization of substorm onsets, even though the details are still debated (see e.g. \citet{Weygand:2008}). 
\begin{figure}
\includegraphics[width=1\textwidth]{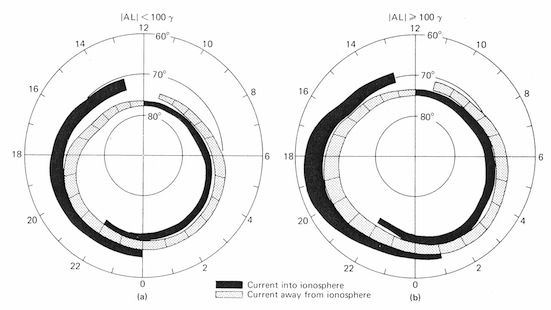}
\caption{Field-aligned current distribution in the polar ionosphere during low geomagnetic activity (left) and during disturbed conditions (right) (1 $\gamma$= 1 nT = $10^{-5}$ G). From \protect\citet{Iijima:1976}, \copyright American Geophysical Union.}
\label{fig:fac}      
\end{figure}

Magnetospheric convection which is driven by reconnection at the dayside and at the tail x-line transports plasma from the tail towards Earth and develops a plasma pressure maximum in the midnight region of the near-Earth tail. The accumulated plasma has to come into equilibrium with the surrounding plasma and the dipole-like magnetic field. That sets up a pressure gradient pointing towards Earth and as the increased pressure wants to relax, it can most easily do so perpendicular to the magnetic field in the east-west direction. This plasma expansion has to overcome friction by ion-neutral collisions in the ionosphere or Ohmic dissipation of the auroral current system which is set up between the magnetosphere and the ionosphere (Figure \ref{fig:bostrom}). The current system is of the Type II as originally suggested by \cite{Bostrom:1964} where the current flows into the ionosphere on the equator-ward side of the auroral arc all the way along the arc and flows out of the ionosphere on the poleward side. This current system is equivalent to the Region-1 and 2 current systems on the evening side and just reversed on the morning side.

\begin{figure}
\includegraphics[width=1\textwidth]{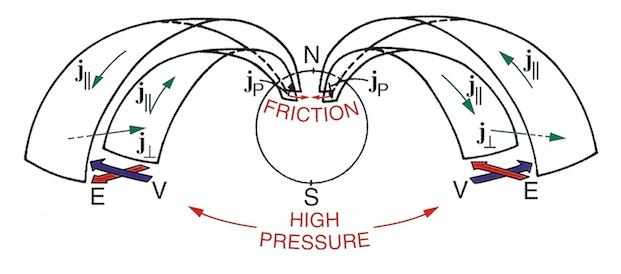}
\caption{Dynamo forces, auroral current system, and resulting convection under frictional control by the ionosphere. From \protect\citet{Paschmann:2003}, \copyright 2002, Kluwer Academic Publishers.}
\label{fig:bostrom}      
\end{figure}

\begin{figure}
\includegraphics[width=1\textwidth]{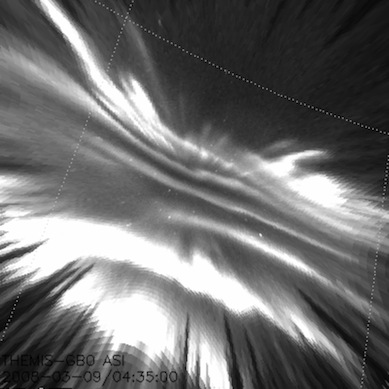}
\caption{Example of an aurora all-sky image mapped onto a geographic grid. The picture was taken on 2008-03-09 at 04:35:00 in Fort Simpson, Canada and shows at least 7 parallel thin auroral arcs. Each side of the mapped image is about 800 km.}
\label{fig:arcs}      
\end{figure}

\paragraph{Magnetic fractures.} 
However, the general picture in Figure \ref{fig:bostrom} is too simple as it depicts a rather static situation, balances only the magnetospheric pressure and ionospheric friction with the energy transfer from the magnetosphere into the ionosphere, and ignores the fact that even during quiet times the aurora very often appears filamented with many parallel arcs stretching over long distances east-west, but having a rather small (1-10 km) north-south extent (Figure \ref{fig:arcs}). Some additional processes must be at play and create the structures and dominate their dynamics that can be seen in the aurora.

In the downward field-aligned current region the current is carried by relatively cold upgoing electrons of ionospheric origin. In the upward field-aligned current region the current is carried by hot downgoing electrons of magnetospheric origin. However, measurements by satellites over auroral arcs showed that the downgoing electrons have much higher energy (2-25 keV) than the electrons in the plasmasheet ($\approx 1$ keV). Therefore an additional acceleration must be at work and early in the space age it was recognized that in the upward current region the parallel current is driven by a field-aligned potential $\Delta \Phi_{\parallel}$ and that it can be related to the physical parameters of the source population through a linear relationship, the Knight relation \citep{Knight:1973}:

\begin{equation}
j_{{\parallel}ion} = K \Delta \Phi_{\parallel} ~~with~~ K=\frac{e^2 n_e}{\sqrt{2 \pi m_e k_B T_e}}
\end{equation}
where K is called the field-aligned or Knight conductance, $T_e$ is the electron temperature, and $n_e$ the density  in the source region. The values for the field-aligned conductance can be deduced from satellite measurements and vary over a rather large range of $10^{-9} - 10^{-11}$ S/m$^2$ ($9 \cdot 10^{-2}-9\cdot 10^{-4}$ (cm s)$^{-1}$). One reason can be that it was deduced ignoring additional particle sources, like trapped electrons, secondary and backscattered electrons \citep{Sakanoi:1995}, and/or wave acceleration \citep{Frey:1998}. The resulting current densities are in the range of several $\mu$A/m$^2$ and current continuity requires that similar current densities are also found in the downward current region. The requirement for current closure combined with the low plasma density along auroral field lines are then the reasons for the generation of substantial potential drops parallel to the magnetic field in the range of a few hundred to a few thousand volts. The current-voltage relation along such magnetic field lines can be determined for simple profiles of the background ion density. For typical parameters, the current density is found to be a few times larger in the downward current region compared to currents in the upward current region for similar potential drops. Thus potential drops up to a few thousand volts and the consequent acceleration of ionospheric electrons up to keV energies, such as has been observed by the FAST satellite, are a necessary consequence of the observed current densities in the downward auroral current region \citep{Temerin:1998}.

\begin{figure}
\includegraphics[width=1\textwidth]{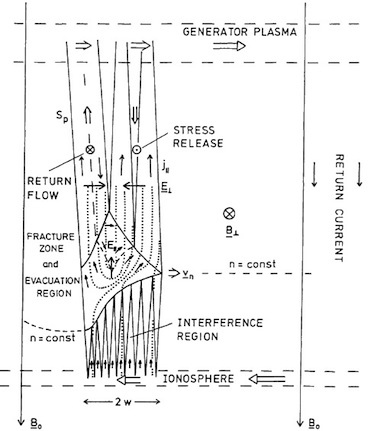}
\caption{Qualitative picture of a magnetic fracture region with the generator region (top) in motion with respect to the ionosphere (bottom). In the fracture region (middle) concentrated field-aligned currents become unstable, allowing finite potential drops and field-aligned acceleration. An interference region between the fracture zone and the ionosphere is characterized by multiple reflections. The fracture zone itself is shown in more detail in Figure \ref{fig:solar_haerendel94}. From \protect\citet{Haerendel:1994}, reproduced by permission of the AAS.}
\label{fig:fracture}      
\end{figure}

Looking at quiet auroral arcs in the evening sector they almost always slowly move equator-ward. As auroral arcs are the ionospheric foot point of the upward field-aligned current the question arises why they are moving and what is the context of their motion. Given the observational fact of motions of arcs and of the oval as a whole it would just be a `logical' conclusion to assume that the whole ionosphere and with it the frozen-in auroral arcs move equator-ward in the evening sector and poleward in the morning sector following the general distorted shape of the auroral oval which extends to much lower geomagnetic latitudes at midnight than it extends at dawn or dusk. Coordinated campaigns of an ionospheric radar measuring the ionospheric plasma motion and an all-sky camera observing the motion of several auroral arcs however, showed that there is a speed difference of 100-200 m/s, and that these arcs perform proper motion \citep{Haerendel:1993, Frey:1996}. Almost always this proper motion is performed into the direction towards the downward field-aligned current.

The relative arc-plasma motion provides a way to supply auroral arcs with both fresh plasma and energy. The process is driven by the `generator' region in the magnetotail which provides the energy that is then used for the conversion into particle kinetic energy and Joule heating $\vec j \cdot \vec E$. Haerendel (1994) studied a system in which an auroral arc propagates into a region of stored magnetic energy, drawing power from a region bounded by upward and downward current sheets. The reason for their propagation is the exhaustion of energy stored in form of magnetic shear stresses in the flux tubes pervading the arc. As these shear stresses are released within time scales of the order of the Alfv\'en transit time, $\tau_A$, between the generator region in the plasma sheet and the auroral acceleration region at 1-2 R$_E$ above the ground, the maximum $(\nabla \times \vec{B})_{\parallel} = 4\pi j_{\parallel}/c$ moves into the stressed field region. Hence the propagation speed must be $v_n = w_{arc}/\tau_A$, where $w_{arc}$ is the effective arc width \citep{Paschmann:2003}.  Some of the consequences of this propagating system are summarized in Figure \ref{fig:fracture}.

The field-aligned current carried from the generator by oblique Alfv\'en waves becomes more intense near the ionosphere because the flux tube converges in the dipole field. At some point this current exceeds a threshold for the generation of micro-instabilities, allowing for a non-zero field-aligned electric field and breakdown of the frozen-in condition. This magnetic `fracture' zone is therefore a region where magnetic field lines break and reconnect again. Most of the liberated magnetic energy is then transferred into kinetic energy of field-aligned accelerated electrons. Most of the energy that is dissipated in the arc is deposited as heat in the ionosphere. A consequence of this heating is the expansion and expulsion of plasma from that region. Most of the ionospheric plasma is excavated by the front of the arc and the so-called auroral density cavity is formed. The fracture zone is highly structured and dynamic, launching disturbances toward the ionosphere in the form of small-scale Alfv\'en waves. These waves can reflect both from the ionosphere and the fracture zone. Multiple reflections lead to highly structured fields in the interference region which can modulate the auroral electron flux and energy. This interference mechanism has been proposed as a potential source of auroral fine structure and multiple arcs as seen in Figure \ref{fig:arcs} \citep{Haerendel:1983}.

\begin{figure}
\includegraphics[width=1\textwidth]{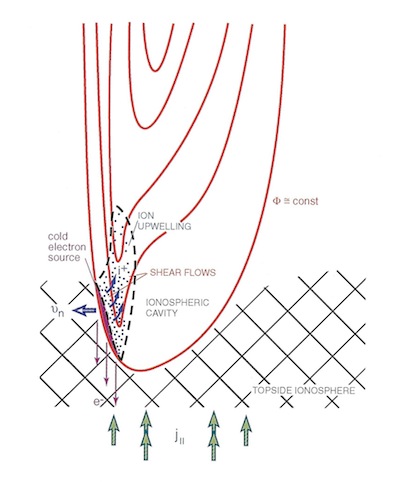}
\caption{Sketch of an auroral cavity. The upward field-aligned current sets up the acceleration region which evacuates the ionospheric plasma by accelerating ions up the field lines and electrons down the field lines and moves into the region with fresh cold electrons. From \protect\citet{Paschmann:2003}, \copyright 2002, Kluwer Academic Publishers.}
\label{fig:cavity}      
\end{figure}

\paragraph{Preventing current starvation.} The evacuation of the plasma and creation of the auroral density cavity has the consequence that in a steady state the auroral current system would undergo a similar effect as the current starvation mentioned earlier in this paper. In addition to extracting energy from the current circuit the proper motion of the auroral arc will maintain the current flow by providing fresh plasma to the acceleration process. To prevent a shortage of current carriers a local field-aligned parallel potential is generated to enlarge the loss cone for thermal magnetospheric electrons. By increasing the parallel velocity of downgoing electrons their pitch angle is decreased and more of them will not mirror in the converging dipole field but will rather penetrate deep enough to be finally lost in the ionosphere through collisions. One clear example where this process happens was demonstrated with High Latitude Dayside Aurora (HiLDA) \citep{Frey:2004}. 

Space-based observations revealed the occurrence of a bright aurora spot well inside the generally empty polar cap. The spot appeared with a 20-60 minutes delay after a substantial drop in solar wind density. No obvious change in solar wind properties could be related to the time of first appearance of the bright spot, which pointed to an internal magnetospheric process rather than a solar wind driven process. Two observations coincided with passes of the FAST satellite just on top of the spots where a strong upward field-aligned current and monoenergetic particles with a clear signature of parallel acceleration were determined. On field lines threading the cusp region (Figure \ref{fig:magnetosphere}) downward field-aligned current was determined \citep{Frey:2003}. Statistical analysis of solar wind conditions demonstrated a clear preference of spot occurrence during positive IMF $B_y$ and $B_z$ when a current circuit is set up \citep{Richmond:1988} with downward field-aligned current in the cusp region, Pedersen current in the ionosphere, and the upward field-aligned current in the polar cap in the center of the dayside convection cell over the HiLDA spot \citep{Le:2002}.

\begin{figure}
\includegraphics[width=1\textwidth]{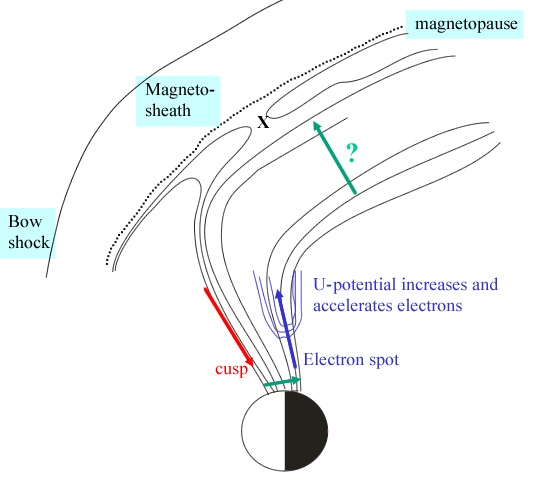}
\caption{Qualitative picture of the field-aligned current distribution for the case of a high latitude dayside aurora spot with the upward current over the spot, downward current connecting the cusp foot point with the high latitude magnetopause, and Pedersen current in the ionosphere. Current closure most likely happens through the magnetopause current.}
\label{fig:hilda}      
\end{figure}

The impact of the low solar wind density on the occurrence of the aurora spot may be twofold. When the solar wind density decreases, the magnetosphere expands in order to maintain pressure balance with the external solar wind plasma medium. This expansion, combined with the reduced supplement of fresh plasma through diffusion, will reduce the number of available current carriers to maintain the upward current. In order to keep the current flowing \citep{Siscoe:2001} the system sets up a parallel potential that accelerates the few existing electrons into the ionosphere and subsequently creates the HiLDA at the footprint, thus avoiding current starvation. The other possible impact of the low solar wind density is the reduction of the Alfv\'en Mach number and a lower plasma $\beta$ in the magnetosheath adjacent to the reconnection site. The reduced Mach number enhances the strength of the dusk convection cell \citep{Crooker:1998} and thus the strength of the upward current at the HiLDA footprint. The lower plasma $\beta$ is considered favorable for the onset of reconnection and can create a stronger downward current into the cusp \citep{Frey:2007}.

\subsection{Conclusions for terrestrial magnetosphere and ionosphere} \label{conclusions_earth}

As mentioned earlier the main difference between electric currents around pulsars or solar flares and currents in the Earth's magnetosphere and ionosphere is that the former are dominated by the effects coming from inside the pulsar or the sun while the latter are dominated and generated by the interaction of the Earth's magnetosphere with the solar wind. The magnetized Earth and the plasma bound by the Earth's magnetic field represent an obstacle to the solar wind which compresses the magnetic field at the dayside and stretches it out into the long tail on the nightside. The magnetopause and neutral sheet currents are driven by this interaction.

Magnetic reconnection transfers plasma and energy from the solar wind into the magnetosphere and drives convection. Magnetic flux is transferred from the dayside magnetosphere to the tail lobes. As the system cannot accumulate energy forever, it has to employ certain methods to get rid of the accumulated energy and one way is to set up field-aligned electric fields that convert electromagnetic energy into kinetic particle energy and drive field-aligned currents. The system also has to remove accumulated plasma and that is achieved through reconnection at the x-line and the ejection of plasmoids down the tail out into the solar wind. It has been shown that the energy conversion above a system of auroral arcs can make a substantial or even larger contribution to the magnetospheric energy loss than ionospheric friction and Ohmic dissipation, and exactly for this reason the notion of an `\emph{auroral pressure valve}' was introduced \citep{Haerendel:2000}. The optical aurora is just a small result of this energy conversion as most of the precipitating particle energy goes into heat and only $\approx 10 \%$ is actually released as energy of photons.

Auroral acceleration is not just found at Earth but on all magnetized planets of the solar system \citep{Bhardwaj:2000, Paschmann:2003}. Additionally, it has also been suggested that auroral acceleration processes happen on other astrophysical objects too, like solar flares, cataclysmic variables, and accreting neutron stars \citep{Haerendel:2001}. Intense field-aligned currents driven by forces in the outer, weak magnetic fields of an astrophysical object are the key to the generation of field-aligned potential drops. In the strong fields near the object the currents may reach a critical limit where the mirror force or current-driven anomalous resistivity require the setup of parallel potential drops for the maintenance of current continuity.

\section{Conclusion}\label{conclusion}
We have investigated electric current systems in three very different magnetized objects in the cosmos: radio pulsar winds, the solar corona, and the terrestrial magnetosphere. Their dimensions and physical conditions differ greatly, yet their main physical effects are very similar.  Essentially, this happens because, in all three cases, an electric current system is set up inside a magnetically dominated plasma by a kinematically dominated plasma, i.e. the rotating magnetic neutron star, the dynamo below the solar surface, and the solar wind. These `drivers' continue to try and increase the free energy of the electric circuit via transport of Poynting flux. However, ultimately, this results in electric current densities which in a number of small spatial domains are too large to be maintained. These domains are the locations of episodic conversion of electric current energy and the release of energized plasma in the form of ejected plasmoids, particle acceleration and heating,  and magnetic turbulence.

These electric circuits are largely force-free inside their magnetically dominated domains, and can be described to a first approximation within the framework of MHD. Yet we underline that it is important to consider the complete electric circuit and not just the force-free part of the magnetic field. While it is well-known that a force-free structure cannot exist on its own and has to be anchored in a non force-free bounding plasma, these boundary conditions are easy to overlook in a magnetic field picture only. When one considers an electric circuit, one automatically includes the voltage source and the resistive regions, which of course are non force-free. Although their relative volumes are minor they do determine the evolution of the system.  

Finally, one would like to know to what extent the end products of bulk motion, particle acceleration and heating, and the release of magnetic turbulence come about by the same physical effects in these different objects. With the proviso that we can make in situ observations inside the terrestrial magnetosphere only, while this is not possible in the corona, let alone in a pulsar wind, we conclude that possibly in all three cases one and the same process does play an important role in converting magnetic/current energy into particle acceleration. This is the occurrence of electric fields along the magnetic field whenever the local plasma conditions cannot provide for the electric current dictated by the global circuit. It is clear, however, that `common' reconnection occurs as well next to or in combination with this `Generalized Magnetic Reconnection'.

\begin{acknowledgements}
HUF was supported through NSF grant AGS-1004736. LF gratefully acknowledges support from the UK's Science and Technology Facilities Council through grant ST/I001808/1, and from the HESPE project (FP7-2010-SPACE-1-263086) funded through the 7th Framework Programme of the European Community. JK thanks the Leids Kerkhoven-Bosscha Fonds for a travel grant. The authors acknowledge the efforts by ISSI for convening the workshop and providing support, which facilitated this joint publication. \end{acknowledgements}

\bibliography{kuijpers-frey-fletcher}

\end{document}